\documentclass[pra,
amssymb,amsmath,amsmath,showpacs,reprint,twocolumn]{revtex4-1}
\usepackage{color}
\usepackage{graphicx}
\usepackage{epstopdf}
\usepackage{natbib}
\usepackage{wasysym}
\usepackage{hyperref}
\usepackage{dcolumn}

\hypersetup{%
   pdfpagemode=None, %FullScreen,
   pdfstartpage=1,
   pdfstartview=FitH,
   pdfmenubar=true,
   pdftoolbar=true,
   colorlinks = true,
   linkcolor=blue,
   citecolor=blue,
   bookmarksopen=false
 }

\newcommand{\Fkt}[1]{\,\mathsf {#1}}

\def\openone{\leavevmode\hbox{\small1\kern-3.3pt\normalsize1}}

\ifx\Tr\renewcommand{\Tr}{\Fkt{Tr}} %has to commented out for IOP
\else\newcommand{\Tr}{\Fkt{Tr}}
\fi

\usepackage{booktabs}
\usepackage{graphicx}
\usepackage{amsmath}
\usepackage{indentfirst}
\begin{document}
\title{Calculation of two-centre two-electron integrals over Slater-type orbitals revisited. \\II.
Neumann expansion of the exchange integrals}

\author{\sc Micha\l\ Lesiuk}
\email{e-mail: lesiuk@tiger.chem.uw.edu.pl}
\author{\sc Robert Moszynski}
\affiliation{\sl Faculty of Chemistry, University of Warsaw\\
Pasteura 1, 02-093 Warsaw, Poland}
\date{\today}

\begin{abstract}
In this paper we consider calculation of two-centre exchange integrals over Slater-type orbitals (STOs). We apply the 
Neumann expansion of the Coulomb interaction potential and 
consider calculation of all basic quantities which appear in the resulting expression. Analytical closed-form equations 
for all auxiliary quantities have already been known but they suffer from large digital erosion when some of the 
parameters are large or small. We derive two differential equations which are obeyed by the most difficult basic 
integrals. Taking them as a starting point, useful series expansions for small parameter values or asymptotic expansions 
for large parameter values are systematically derived. The resulting novel expansions replace the corresponding 
analytical expressions when the latter introduce significant cancellations. Additionally, we reconsider numerical 
integration of some necessary quantities and present a new way to calculate the integrand with a controlled precision. 
All proposed methods are combined to lead to a general, stable algorithm. We perform extensive numerical tests of the 
introduced expressions to verify their validity and usefulness. Advances reported here provide methodology to compute 
two-electron exchange integrals over STOs for a broad range of the nonlinear parameters and large angular momenta.
\end{abstract}

\maketitle

\section{Introduction}
\label{sec:intro}
In the first paper of the series \cite{paper1} (denoted shortly Paper I further in the text), we have discussed
calculation of the two-electron Coulomb and hybrid integrals over
Slater-type orbitals (STOs). The remaining obstacle in the calculations for the diatomic molecules is an accurate
determination of the exchange integrals. These quantities are widely considered to be the most difficult among the
two-centre integrals. There exists a broad literature on this topic with several seminal works written in 50' and 60'
or even earlier. In this introduction let us mention shortly the available methods for the calculation of the exchange
integrals, concentrating solely on the Neumann expansion method which has probably the biggest potential of being
successful for our purposes.

The first comprehensive scheme utilising the Neumann expansion approach was given by Ruedenberg and
co-workers who rewritten the expressions in terms of the charge distributions on both centres
\cite{ruedenberg51,ruedenberg56} and used
a simple one-dimensional numerical quadrature \cite{ruedenberg64a,ruedenberg64b} to avoid analytic integration.
A complete recursive scheme which enabled analytical calculation of all auxiliary quantities along with tabulation of
some integrals was given in the book of Kotani \cite{kotani55}. Short thereafter, Harris \cite{harris60} introduced
significant simplifications in the calculation of some basic integrals by invoking the theory of spherical Bessel
functions. Methodology based on the Neumann expansion was then progressively refined by many authors. Some changes were
introduced which were aimed at improving the efficiency or accuracy of the algorithms and making the final expressions
more transparent and general \cite{yasui82,rico89,rico92,lopez94,rico94}. Importantly, Maslen and Trefry \cite{maslen90}
derived explicit analytical expressions for all basic quantities appearing in the Neumann expansion by using the
hypergeometric function approach. More or less at the same time the limitations of the available expressions were
recognised and a new recursive scheme was proposed by Fern\'{a}ndez Rico and co-workers \cite{rico94b}. More recently,
Harris pursued the approach of Maslen and Trefry based on the analytical expressions, introduced considerable
simplifications and several new expressions which allow more stable calculations of several auxiliary quantities
\cite{harris02}. In this paper we recall some of the works cited above to illuminate the differences and
the advantages of the derived expressions compared to those available in the literature.

The paper is organised as follows. In Sec. II we introduce the notation and present all basic quantities that need to
be calculated. In Sec. III we present analytic expressions available in the literature. We put special emphasis on their
numerical stability and other practical issues. In Sec. IV we derive
a differential equation which is obeyed by the first important function family $L_\mu$. By using it, we introduce two
new methods of calculations for large or small values of the parameters and discuss a general method of evaluation which
is a combination of three analytical methods. Once the problem of the $L_\mu$ functions is solved, in Sec. V we move to 
the calculation of the most important quantities $W_\mu$. We present derivation of the differential equation obeyed
by these functions. Next, new methods for calculation of $W_\mu$ are introduced. In Sec. VI we introduce a seminumerical
method for calculation of the $W_\mu$ functions. We also discuss a general algorithm for calculation of $W_\mu$ 
which is a combination of several known methods. Finally, in Sec. VII we conclude our paper.

In the paper we rely on the known special functions to simplify the derivations and the final formulae.
Our convention for all special functions appearing below is the same as in Ref. \cite{stegun72}.

\section{Preliminaries}
\label{sec:pre}
In this paper, we consider calculation of two-centre exchange integrals in the form
\begin{align}
\label{twoei}
(ab|a'b') = \int d\textbf{r}_1 \int d\textbf{r}_2\, \chi_a^*(\textbf{r}_1) \chi_b(\textbf{r}_1) \frac{1}{r_{12}} 
\chi_{a'}^*(\textbf{r}_2) \chi_{b'}(\textbf{r}_2),
\end{align}
where $r_{12}$ denotes the interelectron distance. For details of the notation see Preliminaries section in Paper I.

The Neumann expansion of $1/r_{12}$ is defined by the following formula \cite{ruedenberg51}
\begin{align}
\begin{split}
\label{neumann}
\frac{1}{r_{12}}&=\frac{2}{R}\sum_{\mu=0}^\infty \sum_{\sigma=-\mu}^\mu (-1)^\sigma (2\mu+1) 
\left[ \frac{(\mu-|\sigma|)!}{(\mu+|\sigma|)!} \right]^2\\
&\times P_\mu^{|\sigma|}(\xi_<)Q_\mu^{|\sigma|}(\xi_>)P_\mu^{|\sigma|}(\eta_1)P_\mu^{|\sigma|}(\eta_2)
e^{\dot{\imath}\sigma(\phi_1-\phi_2)},
\end{split}
\end{align}
where $\xi_<=\min(\xi_1,\xi_2)$ and $\xi_>=\max(\xi_1,\xi_2)$, and other quantities were defined in Paper I. By 
inserting the above expression into Eq. (\ref{twoei}), making use of Eqs. (7), (8) and (9) in Paper I, and carrying out 
the integration over the angles $\phi_1$ and $\phi_2$ one arrives at the formula
\begin{align}
\begin{split}
\label{stoneumann}
(ab|a'b')&=\frac{8}{R}K_{ab}K_{a'b'}\delta_{M_1,-M_2}(-1)^\sigma\\
&\times\sum_{k_1=0}^{k_1^{max}} B_{k_1}^{n_1-l_1,n_2-l_2}\sum_{k_2=0}^{k_2^{max}} B_{k_2}^{n_3-l_3,n_4-l_4} \\
&\times\sum_{p_1,q_1=0}^{\Gamma_1} \left( {\bf \large \Xi}_{l_1l_2}^{M_1} \right)_{p_1q_1}
\sum_{p_2,q_2=0}^{\Gamma_2} \left( {\bf \large \Xi}_{l_3l_4}^{M_2} \right)_{p_2q_2}\\
&\times\sum_{\mu=\sigma}^\infty(2\mu+1)W_\mu^\sigma(p_1+k_1,p_2+k_2,\alpha_1,\alpha_2)\\
&\times i_\mu^\sigma(q_1+k_1^{max}-k_1,\beta_1)i_\mu^\sigma(q_2+k_2^{max}-k_2,\beta_2),
\end{split}
\end{align}
where $\delta_{ij}$ is the Kronecker delta, $K_{ab}$ were defined in Paper I, and $\sigma=|M_1|=|M_2|$ is 
restricted to the regime $\sigma\geq0$. In the above equation a number of new quantities was introduced. We follow the 
notation advocated by Harris \cite{harris02} and others for $\eta$ integration:
\begin{align}
\label{imus}
\begin{split}
i_\mu^\sigma(q,\beta)=\frac{(-1)^\mu}{2}\frac{(\mu-\sigma)!}{(\mu+\sigma)!}
&\int_{-1}^{+1} d\eta \,P_\mu^{|\sigma|}(\eta)\\
&\times(1-\eta^2)^{\sigma/2}\eta^q\, e^{-\beta \eta},
\end{split}
\end{align}
and for $\xi$ integration:
\begin{align}
\begin{split}
\label{bigw}
W_\mu^\sigma(p_1,p_2,\alpha_1,\alpha_2)&=w_\mu^\sigma(p_1,p_2,\alpha_1,\alpha_2)\\
&+w_\mu^\sigma(p_2,p_1,\alpha_2,\alpha_1),\\
\end{split}
\end{align}
\begin{align}
\label{smallw}
\begin{split}
w_\mu^\sigma(p_1,p_2,\alpha_1,\alpha_2)&=\int_1^\infty d\xi_1\, Q_\mu^{\sigma}(\xi_1)\,
(\xi_1^2-1)^{\sigma/2}\xi_1^{p_1}e^{-\alpha_1\xi_1}\\
&\times\int_1^{\xi_1}d\xi_2\, P_\mu^{\sigma}(\xi_2)\,
(\xi_2^2-1)^{\sigma/2}\xi_2^{p_2}e^{-\alpha_2\xi_2}.
\end{split}
\end{align}
The series in Eq. (\ref{stoneumann}) is, in general, infinite. However, it is convergent for
all physically acceptable values of the nonlinear parameters, namely $\alpha_1\geq0$, $\alpha_2\geq0$,
$|\beta_1|\leq\alpha_1$, $|\beta_2|\leq\alpha_2$. The rate of convergence depends critically on the values of $\beta$.
The smaller the $|\beta|$ are, the faster the ellipsoidal expansion converges. In fact, in the limit of $\beta_1=0$ or
$\beta_2=0$ it becomes finite by the virtue of $i_\mu^\sigma(q,\beta)$ integrals which we discuss further. Convergence 
of the Neumann expansion has been studied by several authors and there is a general agreement that at most several tens 
of terms are sufficient to converge all non-negligible integrals to the prescribed accuracy
\cite{yasui82,rico94}. The ``divergence'' of the Neumann expansion reported by other authors \cite{kennedy99} is only 
apparent and results solely from the inaccuracies in calculation of the individual terms.

Let us not discuss the calculation of the $i_\mu^\sigma(q,\beta)$ functions, Eq. (\ref{imus}), which arise from the
integration over $\eta_1$ and $\eta_2$ in Eq. (\ref{stoneumann}). Efficient and accurate recursive formulae that allow
their calculation have been known since the 1960 paper of Harris \cite{harris60}. These expressions were progressively 
refined by many authors \cite{harris02,rico94,rico92,maslen90}. Therefore, we refer to these papers for methods of
computation of $i_\mu^\sigma(q,\beta)$ and consider this problem as entirely solved for the present purposes.

\section{Closed-form analytical formulae for the $\xi_1$, $\xi_2$ integration}
\label{sec:xiint}
The problem of integration over $\xi_1$ and $\xi_2$ \emph{i.e.} accurate calculation of
$W_\mu^\sigma(p_1,p_2,\alpha_1,\alpha_2)$ functions defined in Eq. (\ref{bigw}) is the most important problem in 
practical utilisation of the ellipsoidal expansion. As mentioned in Introduction, explicit analytical expressions for 
these functions were derived by Maslen and Trefry \cite{maslen90}. In this subsection let us recall their formulae and
introduce a number of new quantities. 

Two new important auxiliary functions need to be defined:
\begin{align}
\label{lmuint}
L_\mu^\sigma(p,\alpha)=\frac{(\mu-\sigma)!}{(\mu+\sigma)!}\int_1^\infty d\xi\,
Q_\mu^\sigma(\xi)\xi^p(\xi^2-1)^{\sigma/2} e^{-\alpha \xi},
\end{align}
and
\begin{align}
\label{kmu}
k_\mu^\sigma(p,\alpha)=\frac{(\mu-\sigma)!}{(\mu+\sigma)!}\int_1^\infty d\xi\,
P_\mu^\sigma(\xi)\xi^p(\xi^2-1)^{\sigma/2} e^{-\alpha \xi}.
\end{align}
The following recursion formula was presented by Harris \cite{harris60} and results from the connections of
$k_\mu^\sigma(p,\alpha)$
with the modified spherical Bessel functions:
\begin{align}
k_{\mu+1}^0(0,\alpha) = \frac{2\mu+1}{\alpha}k_\mu^0(0,\alpha)+k_{\mu-1}^0(0,\alpha).
\end{align}
This recurrence requires two starting values: $k_0^0(0,\alpha)=k_{-1}^0(0,\alpha)=e^{-\alpha}/\alpha$ and it is stable 
for all relevant values of $\alpha$ since all terms on r.h.s always have the same sign. Expressions that can be used to 
build $k_\mu^\sigma(p,\alpha)$ with nonzero
values of $p$ and $\sigma$ are derived by inserting appropriate recursion formulae for the Legendre functions
\cite{stegun72}. The working formulae read
\begin{align}
\begin{split}
\label{kmuq}
&k_\mu^\sigma(p+1,\alpha) =\\
&\frac{(\mu-\sigma)k_{\mu-1}^\sigma(p,\alpha)+(\mu+\sigma+1)k_{\mu+1}^\sigma(p,\alpha)}{2\mu+1},
\end{split}
\end{align}
and
\begin{align}
\label{kmusigma}
k_\mu^{\sigma+1}(p,\alpha) = \frac{k_{\mu-1}^\sigma(p,\alpha)-k_{\mu+1}^\sigma(p,\alpha)}{2\mu+1},
\end{align}
and are also completely stable for all acceptable values of $\alpha$. Calculation of $L_\mu^\sigma(p,\alpha)$ is much
more troublesome. The analytical formula was presented by Maslen and Trefry: \cite{maslen90}
\begin{align}
\label{lmugen}
L_\mu^\sigma(p,\alpha) = \sum_s^{\mu+\sigma} \mathcal{A}_s^{\mu\sigma} L_0^0(p+s,\alpha)+\sum_s^{\mu+\sigma-1}
\mathcal{B}_s^{\mu\sigma} A_{p+s}(\alpha),
\end{align}
with
\begin{align}
\label{asmu}
&\mathcal{A}_s^{\mu\sigma} = (-1)^{(\mu+\sigma-s)/2} \frac{(\mu-\sigma+s-1)!!}{s!(\mu+\sigma-s)!!},\\
&\mathcal{B}_s^{\mu\sigma} = \sum_{j=0}^{(\mu+\sigma-s-1)/2}
\frac{(-1)^{j+1}(2\mu-2j-1)!!}{(\mu+\sigma-s-2j)(\mu+\sigma-2j)!(2j)!!}.
\end{align}
where $!!$ denotes the double factorial function. The coefficients $\mathcal{A}_s^{\mu\sigma}$ vanish unless
$\mu+\sigma-s$ is even, and $\mathcal{B}_s^{\mu\sigma}$ are nonzero if and only if $\mu+\sigma-s$ is odd. The quantities
$A_n(\alpha)$ are well-known \cite{mulliken49,corbato56} and were defined in Paper I.

The remaining necessary formula for $L_0^0(p,\alpha)$ can be obtained by direct integration. The result reads
(note the sign error in the original formula \cite{maslen90}):
\begin{align}
\label{l00}
\begin{split}
&L_0^0(p,\alpha)=\frac{1}{2}\Bigg[(-1)^{p+1}E_1(2\alpha)A_p(-\alpha)\\
&+\left[\gamma_E+\log(2\alpha)\right]A_p(\alpha)
+\frac{p!e^{-\alpha}}{\alpha^{p+1}}\sum_{r=1}^{p-1}\mathcal{L}_{pr} \frac{\alpha^r}{r!}
\Bigg],
\end{split}
\end{align}
where $E_1$ is the generalised exponential integral function \cite{stegun72}, $\gamma_E$ is the Euler-Mascheroni
constant and $\mathcal{L}_{pr}$ are numerical coefficients given in the simplified form by Harris \cite{harris02}:
\begin{align}
\mathcal{L}_{pr} = \sum_j^r\left(\frac{1}{j+p-r}-\frac{1}{j}\right)\sum_{k=j}^r (-1)^{r-k}2^k{ r \choose k}.
\end{align}
Once the integrals $L_\mu^\sigma(p,\alpha)$ are calculated the final formula 
for $w_\mu^\sigma(p_1,p_2,\alpha_1,\alpha_2)$, presented by Maslen and Trefry\cite{maslen90} and simplified 
considerably by Harris\cite{harris02}, is as follows:
\begin{align}
\begin{split}
\label{wmugen}
&w_\mu^\sigma(p_1,p_2,\alpha_1,\alpha_2)=\Bigg[ \frac{(\mu+\sigma)!}{(\mu-\sigma)!} \Bigg]^2
\Bigg[ L_\mu^\sigma(p_1,\alpha_1) k_\mu^\sigma(p_2,\alpha_2)\\
&-\sum_s^\mu
\mathcal{A}_s^{\mu\sigma}\sum_{j=0}^{p_2+s}\frac{(p_2+s)!}{j!\alpha_2^{p_2+s}}L_\mu^\sigma(p_1+j,\alpha_1+\alpha_2)
\Bigg]
\end{split}
\end{align}
Since closed-form analytical formulae exist for all basic quantities involved in the
calculation of the integrals, it seems that the problem is solved. This is true, however, only in an
arbitrary arithmetic precision environment such as \textsc{Mathematica} \cite{mathematica7}. Unfortunately, this kind of
environment is too slow to support large scale calculations in the basis sets close to the saturation. For practical 
purposes, we require a theory that allows calculations in a numerically stable way, presumably in the
double or at most quadruple arithmetic precision, for a large range of parameter values and high angular momenta. 

Reasons for the observed numerical instabilities were partially recognised by Maslen and Trefry \cite{maslen90} and
analysed in details by Harris \cite{harris02}. The latter paper should be consulted for a more elaborate study of the 
numerical instabilities. We give only a superficial overview of the problem.

Two the most important working formulae, Eqs. (\ref{lmugen}) and (\ref{wmugen}), are both
numerically badly conditioned. Equation (\ref{lmugen}) is unstable for small values of $\alpha$ when 
$\mu$ and/or $p$ are moderate or large. This happens due to large cancellation occurring between the first and the 
second term, which are both large, nearly equal and of opposite signs. Our numerical tests have shown that this formula
provides a sufficient level of accuracy for all practically required $\mu$, $p$ and $\sigma$ only if $\alpha\geq3$. This
agrees more or less with the conclusion of Maslen and Trefry. Unfortunately, typical basis sets give rise to the
integrals which require $\alpha$ to be considerably smaller than that. Therefore, Eq. (\ref{lmugen}) alone is not
sufficient to calculate all necessary integrals with a controlled precision. Thus, a different method has to be devised
for the small $\alpha$ regime.

A similar situation is encountered in Eq. (\ref{wmugen}). The cancellation occurs between two terms in the second
square bracket when the value of $\alpha_2$ is small. However, when accurate values of $L_\mu^\sigma(p,\alpha)$ are
provided at start, good accuracy is retained for all reasonable values of $\alpha_1$, if only $\alpha_2$ is moderate or
large. Of course, instabilities become increasingly severe for higher values of $\mu$, $p_2$ and $\sigma$. Similarly as
in the previous case, the value of $\alpha_2$ need to be large enough to make Eq. (\ref{wmugen}) useful. Above
$\alpha_2=3$ the instabilities are not severe for all $\mu$, $p_1$, $p_2$ and $\sigma$ needed in practice. In such
case Eq. (\ref{wmugen}) provides a reasonable way to build the necessary $w_\mu^\sigma(p_1,p_2,\alpha_1,\alpha_2)$, if 
only
sufficiently accurate values of $L_\mu^\sigma(p,\alpha)$ are available. For smaller values of $\alpha_2$ a different
method is required.

Other quantities entering Eq. (\ref{stoneumann}) do not pose significant numerical difficulties during evaluation by
using closed-form expressions. Similarly, rather a minor loss of digits is observed during summation of the Neumann
expansion or post-processing of the resulting integrals. Therefore, instabilities connected with Eqs. (\ref{lmugen})
and (\ref{wmugen}) are the main obstacles to accurate calculation of two-centre two-electron exchange integrals over
Slater-type
orbitals by using the ellipsoidal expansion.

\section{Calculation of the $L_\mu^\sigma(p,\alpha)$ functions }
\label{sec:lmu}

Before presenting the working formulae let us recall two simple recursions which allow to 
simplify the problem considerably. Both of them were recognised in the early works \cite{ruedenberg51,ruedenberg56} and 
result directly from the
properties of the Legendre functions and the integral representation (\ref{lmuint}). They are as follows:
\begin{align}
\label{lgrowp}
L_\mu^0(p+1,\alpha) &= \frac{(\mu+1)L_{\mu+1}^0(p,\alpha)+\mu L_{\mu-1}^0(p,\alpha)}{2\mu+1},\\
\label{lgrows}
L_\mu^{\sigma+1}(p,\alpha) &= \frac{L_{\mu+1}^{\sigma}(p,\alpha)-L_{\mu-1}^{\sigma}(p,\alpha)}{2\mu+1}.
\end{align}
The first of the above equations is completely stable when used as it stands \emph{i.e.} in the direction of increasing
$p$. Conversely, it is very unstable when used to increase $\mu$ at the cost of $p$. The only disadvantage of this 
formula
is that one has to build $L_\mu^0(0,\alpha)$ with much higher $\mu$ than normally needed in order to produce the
required $L_\mu^0(p,\alpha)$ integrals up to a given $p_{max}$. 
% Harris proposed to supplement this formula with an
% another expression that allows to reduce the required level of $\mu$ considerably but at cost of loosing some digits
% \cite{harris02}. However, in our approach the necessity to calculate $L_\mu^0(0,\alpha)$ integrals, even with a quite
% large values of $\mu$, does not pose a difficulty. Therefore, it seems optimal to use Eq. (\ref{lgrowp}) by default. 

Considering Eq. (\ref{lgrows}), it can be used to build the final integrals $L_\mu^{\sigma}(p,\alpha)$ starting from the
integrals with $\sigma=0$. Most importantly, this recursion is also numerically stable for all relevant values of
$\alpha$ when used for this purpose. To sum up, Eqs. (\ref{lgrowp}) and (\ref{lgrows}) allow us to build
$L_\mu^{\sigma}(p,\alpha)$ integrals very accurately, if only sufficiently correct values of $L_\mu^0(0,\alpha)$ are
provided. Obviously, since now we require very large values of $\mu$ ($\mu=50$ is not an overestimation),  Eq. 
(\ref{lmugen}) cannot be used for that purpose for small values of $\alpha$. 

Before passing further, let us introduce an abbreviation $L_\mu(\alpha):=L_\mu^0(0,\alpha)$ which allows to make our
equations more compact and transparent.

\subsection{Recursive calculation of the $L_\mu(\alpha)$ functions }
\label{subsec:relmu}
Let us first note that Eqs. (\ref{lmugen}) and (\ref{l00}) are not perfectly suitable for a numerical evaluation.
Despite their apparent simplicity, they introduce a number of auxiliary quantities which require a separate calculation.
The optimal strategy is to utilise a recursive formula that connects the values of $L_\mu^0(0,\alpha)$ with
different $\mu$, since all integrals up to a given $\mu_{max}$ are used to evaluate a bunch of the two-electron
integrals. The required formula was obtained by Harris \cite{harris02}
\begin{align}
\label{lmuhar}
L_{\mu+1}(\alpha)-\frac{2\mu+1}{\alpha}L_\mu(\alpha)-L_{\mu-1}(\alpha)=
-\frac{2\mu+1}{\mu(\mu+1)}\frac{e^{-\alpha}}{\alpha}.
\end{align}
Unfortunately, this formula is not free from the numerical instabilities. The upward recursion is
hopelessly unstable and probably useless. The downward recursion is also unstable, but in a more gentle and subtle way.
As
observed by Harris, the instabilities in the downward recursion arise only from contamination of the initial values by
a multiple of the solution to the complementary recurrence problem \emph{i.e.} formula (\ref{lmuhar}) with
the right hand side neglected. Solutions of the complementary recurrence problem are $(-1)^\mu i_\mu(\alpha)$ and
$(-1)^\mu k_\mu(\alpha)$, multiplied by an $\alpha$-dependent factor. The latter solution vanishes quickly and does not
contaminate the final values. Threfore, the former is responsible for the numerical instabilities. As a result, the 
following
algorithm can be proposed (downward recursion with adjustment):
\begin{enumerate}
\item start with the exact initial values of $L_{\mu_{max}}(\alpha)$ and $L_{\mu_{max-1}}(\alpha)$, and carry out the
recursive process (\ref{lmuhar}) downward until $\mu=0$ is reached,
\item calculate $i_\mu(\alpha)$ for $\mu$ up to $\mu_{max}$,
\item compute $L_0(\alpha)$ by using the formula (\ref{l00}) and find the weighted difference
$F=\frac{L_0(\alpha)-\bar{L}_0(\alpha)}{i_0(\alpha)}$, where $\bar{L}_\mu(\alpha)$ denotes the results of the downward
recurrence,
\item perform the ``adjustment'': $L_\mu(\alpha) = \bar{L}_\mu(\alpha) + F\times (-1)^\mu i_\mu(\alpha)$.
\end{enumerate}
In the last step of the algorithm the error resulting from the contamination is eliminated. This is, in fact, a 
special case of the back substitution known from the Olver algorithm \cite{olver67,olver72}. There are two
major difficulties connected with the above procedure. Firstly, one requires accurate values of
$L_{\mu_{max}}(\alpha)$ and $L_{\mu_{max-1}}(\alpha)$ to be provided at start. Secondly, the downward recursion has to 
be restarted occasionally with ``fresh'' values of $L_\mu(\alpha)$,
somewhere in the middle of the recursive process to keep the value of $F$ reasonably small.
% , before $\mu=0$ is reached. Without restarting, contamination error
% grows very fast for some values of $\alpha$. Thus, unacceptably large loss of digits may occur during computation of
% $F$.

The second problem can be solved in a brute-force fashion. In our implementation the downward recursion is restarted
after each ten steps. For instance, when $\mu_{max}=30$ is required, the restarts occur at $\mu=20$ and $\mu=10$. 
% We found it sufficient to preserve a reasonable accuracy for all $\mu$ during the ``adjustment'' process. 

Much more troublesome problem is the calculation of the initial values, $L_{\mu_{max}}(\alpha)$ and
$L_{\mu_{max-1}}(\alpha)$, for a given $\mu_{max}$. We believe that the idea of Harris was to calculate those values
from the explicit expressions by using an extended arithmetic precision. However, this requires an arbitrary precision 
package to be available and can be very time consuming,
especially when a large number of restarts is required. We propose a different approach which is closely related to the
methods of computation of the Boys function \cite{boys50} in the Gaussian integral theories \cite{gill91}. Namely, for
all required $\mu$ we created
a look-up tables which contain the values of the function $L_\mu(\alpha)$ and its several derivatives with respect to
$\alpha$, calculated on a properly suited grid. The grid spans over the interval $\alpha\in[0,100]$ with the step 
$0.01$. We tabulated the data for $L_\mu(\alpha)$ with $\mu=10,20,...,100$, which is sufficient for the practical
purposes.
Once the described look-up tables are created, the calculation of $L_\mu(\alpha)$ goes as follows. At input, the desired
value of $\alpha$ is provided. If this value hits exactly one of the grid points then the final value of $L_\mu(\alpha)$
is immediately returned. If not, the closest grid point is found and the values of $L_\mu(\alpha)$ and its several
derivatives at this point are read from the look-up tables. Then, the Taylor expansion around the chosen grid point is
performed which allows to compute the value of $L_\mu(\alpha)$ for the desired $\alpha$. Even a better performance of
this approach can be reached if an expansion in the Chebyshev polynomials \cite{stegun72} is used instead of the
ordinary Taylor series.
However, our tests showed that the gain is fairly minor for this specific problem.

% The presented method requires a lot of effort during creation of the look-up tables since they need to be computed in an
% extended arithmetic precision and carefully checked. However, once the method is implemented, 
It is obvious that the method based on the look-up tables is stable and inexpensive. However, two usual problems are 
connected with this approach. Firstly, the Taylor
expansion tends to break down once $\alpha$ is close to zero. Secondly, only a finite look-up tables can be stored, so
that calculations are supported only up to some large value of $\alpha$. In the following subsections, we shall present 
two methods which are perfectly suitable for calculations of $L_\mu(\alpha)$ in either small or large $\alpha$ regime.

\subsection{The differential equation for the $L_\mu(\alpha)$ functions }
\label{sbsec:diflmu}
In this subsection we present a different approach to calculation of the $L_\mu^\sigma(p,\alpha)$ functions. Most of the
working formulae available in the literature were derived starting from the integral representation, Eq.
(\ref{lmuint}). We propose another line of attack, to derive a differential equation with respect to $\alpha$ which
is obeyed by these integrals. The obtained differential equation is then solved by using properly tailored series
expansions, substituting available analytical expressions in the regions where they lose numerical stability.

The derivation of the differential equation for $L_\mu(\alpha)$ starts with the well-known expression:
\begin{align}
\label{eqq}
(1-\xi^2)Q_\mu''(\xi)-2\xi Q_\mu'(\xi)+\mu(\mu+1)Q_\mu(\xi)=0,
\end{align}
where $Q_\mu$ are the Legendre functions of the second kind. By multiplying the above expression by
$e^{-\alpha \xi}$ and integrating over $\xi$ on the interval $[1,+\infty]$ one obtains:
\begin{align}
\begin{split}
0&=\int_1^\infty Q_\mu''(\xi) e^{-\alpha \xi}-\int_1^\infty Q_\mu''(\xi) \xi^2e^{-\alpha \xi}\\
&-2\int_1^\infty Q_\mu'(\xi) \xi e^{-\alpha \xi}+\mu(\mu+1)\int_1^\infty Q_\mu(\xi) e^{-\alpha \xi}.
\end{split}
\end{align}
The first two integrals need now to be integrated by parts twice in order to get rid of the derivatives of the Legendre
functions. The results are:
\begin{align}
\begin{split}
\int_1^\infty Q_\mu''(\xi) e^{-\alpha \xi}&=
-e^{-\alpha}\lim_{\xi\rightarrow1^+}\left[Q_\mu'(\xi)+\alpha Q_\mu(\xi)\right]\\
&+\alpha^2 L_\mu(\alpha),\\
\int_1^\infty Q_\mu''(\xi) \xi^2e^{-\alpha \xi}&=
-e^{-\alpha}\lim_{\xi\rightarrow1^+}\xi^2\left[Q_\mu'(\xi)+\alpha Q_\mu(\xi)\right]\\
&-2\int_1^\infty Q_\mu'(\xi) \xi e^{-\alpha \xi}
-2\alpha L_\mu^0(1,\alpha)\\
&+\alpha^2 L_\mu^0(2,\alpha).
\end{split}
\end{align}
By inserting the above expressions into the initial equation (\ref{eqq}) we see that the third integral is
cancelled out. Additionally, it follows trivially from the definition that $L_\mu'(\alpha):=\frac{\partial}{\partial
\alpha}L_\mu(\alpha)=-L_\mu^0(1,\alpha)$ and $L_\mu''(\alpha):=\frac{\partial^2}{\partial
\alpha^2}L_\mu(\alpha)=L_\mu^0(2,\alpha)$. By making proper rearrangements one obtains readily
\begin{align}
\label{lmudiff1}
\begin{split}
0&=e^{-\alpha} \lim_{\xi\rightarrow1^+} (\xi^2-1)Q_\mu'(\xi)
+\alpha\, e^{-\alpha} \lim_{\xi\rightarrow1^+} (\xi^2-1)Q_\mu(\xi)\\
&-\alpha^2 L_\mu''(\alpha)-2\alpha L_\mu'(\alpha)+\left[\mu(\mu+1)+\alpha^2\right]L_\mu(\alpha). 
\end{split}
\end{align}
To calculate the limits in the above expression we have to recall the following series expansion of
the Legendre functions of the second kind around the point $\xi=1^+$:
\begin{align}
\begin{split}
Q_\mu(\xi)&=\frac{1}{2}\log(2)-\frac{1}{2}\log(\xi-1)\\
&-\gamma_E-\psi(\mu+1)+\mathcal{O}(\xi-1),
\end{split}
\end{align}
where $\psi(z)$ is the digamma function \cite{stegun72}, so that the above limits evaluate to
\begin{align}
&\lim_{\xi\rightarrow1^+} (\xi^2-1)Q_\mu'(\xi)=-1,\\
&\lim_{\xi\rightarrow1^+} (\xi^2-1)Q_\mu(\xi)=0.
\end{align}
Therefore, the final form of the differential equation is disarmingly simple:
\begin{align}
\label{lmudiff2}
\alpha^2 L_\mu''(\alpha)+2\alpha L_\mu'(\alpha)-\left[\mu(\mu+1)+\alpha^2\right]L_\mu(\alpha)=-e^{-\alpha}.
\end{align}
% This differential equation confirms the close connection between the $L_\mu^0(0,\alpha)$ and $k_\mu^0(0,\alpha)$
% functions. When the Legendre functions $Q_\mu$ are replaced by $P_\mu$ in the integral representation (\ref{lmuint})
% then both limits present in Eq. (\ref{lmudiff1}) evaluate to zero since the $P_\mu$ functions are regular around the
% point $\xi=1$. As a result, the exponential term on the right hand side of Eq. (\ref{lmudiff2}) vanishes and the
% differential equation becomes exactly the same as for the modified spherical Bessel functions $k_\mu$. This gives an
% alternative method to recognise $k_\mu^0(0,\alpha)$ in terms of the Bessel functions which was originally done by
% invoking a proper integral representation.

We believe that the differential equation (\ref{lmudiff2}) has not been known to the previous investigators. It is
not obvious at first glance, however, if it can be used in practice. The next two subsections of this paper are devoted
to the derivation of the solutions of this differential equation which are applicable either in small or large
$\alpha$ regimes. Especially the small $\alpha$ region is of primary importance since it is the regime where both the
analytical expression, Eq. (\ref{lmugen}), and the recursive method, Eq. (\ref{lmuhar}), fail to provide accurate
results.

\subsection{Calculation of the $L_\mu(\alpha)$ functions for small values of $\alpha$}
\label{subsec:lmusmall}
Considering the differential equation (\ref{lmudiff2}) it is trivial to conclude that the general solution can be
written as:
\begin{align}
\label{lmug}
C_\mu^{(1)}i_\mu(\alpha)+C_\mu^{(2)}k_\mu(\alpha)+\mathcal{L}_\mu(\alpha),
\end{align}
where $i_\mu(\alpha):=i_\mu^0(0,\alpha)$ and $k_\mu(\alpha):=k_\mu^0(0,\alpha)$ are the solutions of the homogeneous
problem, modified spherical Bessel functions. The constants $C_\mu^{(i)}$, $i=1,2$, depend solely on $\mu$
and can be fixed thereafter by imposing proper initial conditions. The main problem lies then in the determination of
the particular solution $\mathcal{L}_\mu(\alpha)$. Since in this subsection we are interested only in providing a new
method applicable in the small $\alpha$ regime, our guess for the particular solution is
\begin{align}
\label{lmug2}
\mathcal{L}_\mu(\alpha) = \sum_{k=0}^\infty c_k^\mu \alpha^k + \log(2\alpha)\sum_{k=\mu(+2)}^\infty d_k^\mu \alpha^k,
\end{align}
where the subscript $k=\mu(+2)$ denotes that the summation over $k$ starts at $\mu$ and runs with the step of $2$. In
other words, we assume further that $d_{\mu+1}^\mu$, $d_{\mu+3}^\mu$,... are zero. The above formula is not
intuitive and should to some extent be reasoned. The presence of the logarithmic terms is necessary by the virtue of
the formula (\ref{lmugen}) (note the zero limit). The summation range in the second term is chosen \emph{ad hoc} to 
give the
simplest possible starting point that we are aware of. We could guess blindly that the second summation starts from zero
and runs with the unit step. However, this shall make the derivation much more tedious without changing the
final conclusion.

By inserting the above formula into the differential equation (\ref{lmudiff2}) and grouping the same powers of $\alpha$
together one obtains expression for the terms proportional to $\log(2\alpha)$:
\begin{align}
\sum_{k=\mu+2(+2)}^\infty \alpha^k \big[ k(k+1)d_k^\mu-\mu(\mu+1)d_k^\mu-d_{k-2}^\mu\big]=0,
\end{align}
which leads to a well-defined recursion relation
\begin{align}
\label{recds}
d_k^\mu = \frac{d_{k-2}^\mu}{k(k+1)-\mu(\mu+1)},\;\;\;\;\mbox{for $k>\mu$},
\end{align}
with $d_\mu^\mu$ being arbitrary for a moment. Powers of $\alpha$ which are not multiplied by the logarithmic terms
split
naturally into two groups, giving rise to the second and the third expression:
%\begin{widetext}
\begin{align}
\begin{split}
\label{recc1}
&\sum_{k=2}^{\mu-1}\alpha^k\Bigg[c_k^\mu
k(k+1)-\mu(\mu+1)c_k^\mu-c_{k-2}^\mu
+\frac{(-1)^k}{k!}\Bigg]=0,
\end{split}
\end{align}
\begin{align}
\begin{split}
\label{recc2}
&\sum_{k=\mu}^\infty \alpha^k \Bigg[c_k^\mu k(k+1)+d_k^\mu(2k+1)-\mu(\mu+1)c_k^\mu \\
&-c_{k-2}^\mu+\frac{(-1)^k}{k!}\Bigg]=0.
\end{split}
\end{align}
%\end{widetext}
Additionally, one obtains two initial values of the coefficients $c_0^\mu=\frac{1}{\mu(\mu+1)}$ and
$c_1^\mu=\frac{1}{2-\mu(\mu+1)}$ (from the indicial equation). Equations (\ref{recc1}) and (\ref{recc2}) can be solved 
to
give the following recursion relations:
\begin{align}
\label{recc1s}
c_k^\mu k(k+1)-\mu(\mu+1)c_k^\mu-c_{k-2}^\mu+(-1)^k/k!=0,
\end{align}
for $k\leq\mu-1$, and
\begin{align}
\begin{split}
\label{recc2s}
&c_k^\mu k(k+1)+d_k^\mu(2k+1)-\mu(\mu+1)c_k^\mu\\
&-c_{k-2}^\mu+(-1)^k/k!=0,
\end{split}
\end{align}
for $k\geq\mu+1$. The value of $d_\mu^\mu$ is fixed by the relation:
\begin{align}
\label{recd2}
d_\mu^\mu=\frac{c_{\mu-2}^\mu+(-1)^{\mu+1}/\mu!}{2\mu+1}.
\end{align}
Therefore, all coefficients appearing in the expression (\ref{lmug2}) are fixed by proper recursion relations apart
from one remaining coefficient $c_\mu^\mu$ which can be fixed arbitrarily. The choice of this coefficient is purely
conventional since it works in a tandem with the constants $C_\mu^{(i)}$, $i=1,2$, from Eq. (\ref{lmug}). For
instance, one can put $C_\mu^{(1)}$ equal to zero and then fix the value of $c_\mu^\mu$ from the initial conditions.
Equivalently, $c_\mu^\mu$ can be zeroed and the values of $C_\mu^{(i)}$  used to meet the initial conditions.
Any intermediate choice is also acceptable. For simplicity and numerical convenience we put $c_\mu^\mu=0$
and transfer the responsibility for fulfilling the initial conditions to the coefficients $C_\mu^{(i)}$. 

The equations presented here are rather simple, linear recursions which can be implemented 
efficiently. It is also easy to verify, even numerically, that no loss of digits
is observed when these recursions are carried out. The only complication is that one can get lost
in the order in which the recursions need to be performed. To address this problem, let us present a sketch of the
algorithm that allows calculation of $\mathcal{L}_\mu(\alpha)$ for a given $\mu$ and $\alpha$:
\begin{enumerate}
\item calculate $c_0^\mu=\frac{1}{\mu(\mu+1)}$ and $c_1^\mu=\frac{1}{2-\mu(\mu+1)}$,
\item build $c_k^\mu$ for $k\leq\mu-1$ by using Eq. (\ref{recc1s}),
\item calculate $d_\mu^\mu$ from Eq. (\ref{recd2}),
\item build $d_k^\mu$ for $k\geq\mu+2$ by using Eq. (\ref{recds}) up to a predefined value $k_{max}$,
\item set $c_\mu^\mu=0$ (see text for the discussion),
\item build $c_k^\mu$ for $k\geq\mu+1$ by using Eq. (\ref{recc2s}) up to a predefined value $k_{max}$,
\item perform the final summations according to Eq. (\ref{lmug2}) and break them off at $k_{max}$.
\end{enumerate}
Alternatively, the values of $c_k^\mu$ or $d_k^\mu$ can be tabulated. 

Having said this, the only remaining issue is the determination of the constants $C_\mu^{(i)}$, $i=1,2$. The second
coefficient can be fixed almost immediately from the following reasoning. The functions $L_\mu(\alpha)$ possess at most
logarithmic singularities as $\alpha$ tends to zero. On the other hand, $k_\mu(\alpha)$ functions possess strong
$1/\alpha^\mu$ type singularities in the limit of small $\alpha$. Since $L_\mu(\alpha)$ cannot be contaminated by
these kind of singularities, the constant $C_\mu^{(2)}$ must be fixed to zero for every $\mu$. The remaining problem
is to give an analytical expression for the constants $C_\mu^{(1)}$. The derivation flows without difficulties
but it is rather tedious. Therefore, it has been moved to Supplemental Material of this paper \cite{supplement}. 
Obviously, it is easy to 
predict in advance how large values of $\mu$ are necessary in practice. Therefore, the values of $C_\mu^{(1)}$ can be 
included in a production program as a simple look-up table. This helps to minimise the cost of the corresponding 
calculations. 

% We need to add that the coefficient $C_\mu^{(1)}$ has to be considered as a function of a single discrete variable
% $\mu$. Since it is easy to predict in advance how large values of $\mu$ are necessary in practice, it seems quite
% pointless to include the above expression into the program and evaluate it every time the corresponding subroutine is
% called. Much simpler approach is to calculate $C_\mu^{(1)}$ up to some large value of $\mu$ only
% once, presumably in a higher arithmetic precision. The resulting data can be then included in a production program in
% form of a simple look-up table. This helps to minimise the cost of the corresponding calculations.

We have to note that an alternative approach to the calculation of $L_\mu^\sigma(p,\alpha)$ functions for small $\alpha$
exists. It was presented by Maslen and Trefry and significantly reformulated by Harris. Their working formula reads:
\begin{align}
\label{lmuh}
\begin{split}
L_\mu^\sigma(p,\alpha) &= \sum_{k=0(+2)}^{p-\mu+\sigma-1}
\frac{(-1)^\sigma(k+\mu-\sigma)!}{(k+2\mu+1)!!k!!}A_{p-\mu+\sigma-k-1}(\alpha)\\
&+(-1)^\mu i_\mu^\sigma(p,\alpha)E_1(\alpha)+e^{-\alpha}\sum_{t=0}^\infty \mathcal{M}_t^{\mu\sigma}(p)\alpha^t,
\end{split}
\end{align}
where
\begin{align}
\begin{split}
\label{lmuh2}
\mathcal{M}_t^{\mu\sigma}(p) &= \sum_{j=0}^{(\mu+\sigma)/2}\sum_{l=0}^t
\frac{(-1)^{l+j}(2\mu-2j-1)!!}{(\mu+\sigma-2j)!l!(t-l)!(2j)!!}\\
&\times T(2k_1+2\mu-2j+1,2k_1+\mu-\sigma-p-l),
\end{split}
\end{align}
and $T(i,j)$ stands for the summation $\sum_{k=0}^\infty 1/(2k+i)(2k+j)$.
% which can be brought into a closed form depending on the parity of $i$ and $j$. We have implemented this method to 
% compare its behaviour with the one proposed in this subsection. Generally speaking, our method has a slightly better 
% convergence rate of the infinite summations, accuracy and range of the applicability. 
The above expressions need to be strictly equivalent to the new method presented by us. Therefore, the only 
difference between the method of Harris and ours lies in the computational costs. The former requires 
calculations of the 
coefficients $\mathcal{M}_t^{\mu\sigma}(p)$ which are rather complicated, four-index quantities. Since it is not known 
in 
advance what is the biggest value of $t$ required, it becomes risky to tabulate these coefficients. Production of 
the corresponding look-up table is cumbersome anyway. On the other hand, no recursive formulae that connect 
$\mathcal{M}_t^{\mu\sigma}(p)$ with different values of the parameters are known. Our method consists only of carrying 
out several simple linear recursions. Moreover, it is not necessary to evaluate any special functions such as $E_1$. To 
sum up, we found that our method is at least one order of magnitude faster for typical values of $\mu$. Taking into 
consideration that both methods perform similarly when it comes to the accuracy, we recommend to use our method 
throughout.

To verify the validity of the new method we have benchmarked it against the results of calculations using the 
closed-form expressions. In order to obtain the reference values, we used the extended arithmetic precision provided 
by the
\textsc{Mathematica} package. Results of the representative calculations are shown in Table \ref{table1}. We see that in
the region $\alpha \leq 1$ our method behaves excellently. Additionally, one can verify that the closed-form formula, 
Eq. (\ref{lmugen}), fails to provide accurate values for small $\alpha$. With our new method it is probably even 
possible to get accurate results for somewhat bigger values of $\alpha$, at cost of including some extra terms in Eq.
(\ref{lmug2}). However, above $\alpha=1$ the recursive method (Subsection \ref{subsec:relmu}) kicks in and it is 
favoured in this regime.

Finally, let us mention that the equations presented here are also valid for $\mu=0$. The main working equations 
(\ref{recds}) and (\ref{recc2s}) remain valid and generate no singularities, but become considerably simplified. For
instance, recurrence relation (\ref{recds}) can now be explicitly solved to give $d_k=-1/(k+1)!$. To be consistent with
the adopted convention, one has to put $c_0^0=0$ and $c_1^0=1/2$. As an equivalent of Eq. (\ref{recd2}) we have
$d_0^0=-1$ and Eq. (\ref{recc1s}) is of no use for $\mu=0$. Additionally, to match the convention adopted in
Supplemental Material \cite{supplement}, as the solutions of the homogeneous differential equation one picks up $\sinh 
(\alpha)/\alpha$ 
and $\cosh
(\alpha)/\alpha$. The latter solution can then be neglected due to strong singularity at the origin. As presented in 
Table \ref{table1}, numerical stability of the method for $\mu=0$ is also very good.

\begin{table*}[ht]
\caption{Exemplary calculations of the $L_\mu^0(0,\alpha)$ functions for some representative values of $\alpha$. 
\emph{Exact}
denotes values calculated using Eq. (\ref{lmugen}) in extended arithmetic precision of 120 significant digits with the
\textsc{Mathematica} package (all digits shown are correct). \emph{Closed-form} denotes calculations with Eq.
(\ref{lmugen}) in the double precision arithmetic (around 15 significant figures). \emph{New} column shows results of
calculations with Eqs. (\ref{lmug}) and (\ref{lmug2}), also in the double precision arithmetic. \emph{Convergence}
denotes a number of terms in Eq. (\ref{lmug2}) required to converge both summations to 15 significant digits. The symbol
$[k]$ denotes the powers of 10, 10$^k$.}
\begin{ruledtabular}
\begin{tabular}{ccccc}
\label{table1}
$\mu$ & exact & closed-form & new & convergence \\
\hline\\[-2.1ex]
\multicolumn{5}{c}{ $\alpha=0.1$} \\
\hline\\[-2.1ex]
0  & 2.08 622 255 552 379 [$+$00] & 2.08 622 255 552 380 [$+$00] & 2.08 622 255 552 379 [$+$00] & 9\\
5  & 2.99 492 885 109 320 [$-$02] & 2.99 496 650 695 801 [$-$02] & 2.99 492 885 109 320 [$-$02] & 9\\
10 & 8.21 061 998 897 917 [$-$03] & 2.43 200 000 000 000 [$+$04] & 8.21 061 998 897 917 [$-$03] & 9\\
15 & 3.76 699 311 176 390 [$-$03] & 3.41 288 409 261 670 [$+$16] & 3.76 699 311 176 390 [$-$03] & 9\\
20 & 2.15 334 499 798 711 [$-$03] & 1.49 481 259 743 710 [$+$29] & 2.15 334 499 798 711 [$-$03] & 8\\
25 & 1.39 162 818 542 327 [$-$03] & 2.43 369 948 821 855 [$+$42] & 1.39 162 818 542 327 [$-$03] & 8\\
30 & 9.72 733 864 877 070 [$-$04] & 9.69 303 429 597 675 [$+$55] & 9.72 733 864 877 070 [$-$04] & 8\\[0.5ex]
\hline\\[-2.1ex]
\multicolumn{5}{c}{ $\alpha=1.0$} \\
\hline\\[-2.1ex]
0  & 3.00 132 871 666 711 [$-$01] & 3.00 132 871 666 711 [$-$01] & 3.00 132 871 666 711 [$-$01] & 20\\
5  & 1.15 009 425 728 751 [$-$02] & 1.15 009 425 727 806 [$-$02] & 1.15 009 425 728 751 [$-$02] & 19\\
10 & 3.28 467 374 818 315 [$-$03] & 3.28 468 531 370 163 [$-$03] & 3.28 467 374 818 315 [$-$03] & 19\\
15 & 1.52 016 466 579 821 [$-$03] & 4.37 500 000 000 000 [$+$01] & 1.52 016 466 579 821 [$-$03] & 19\\
20 & 8.71 752 414 027 890 [$-$04] & 4.82 344 960 000 000 [$+$07] & 8.71 752 414 027 890 [$-$04] & 18\\
25 & 5.64 232 304 101 147 [$-$04] & 2.25 179 981 368 525 [$+$15] & 5.64 232 304 101 147 [$-$04] & 18\\
30 & 3.94 720 438 208 518 [$-$04] & 3.92 900 891 374 755 [$+$24] & 3.94 720 438 208 517 [$-$04] & 18\\
\end{tabular}
\end{ruledtabular}
\end{table*}

\subsection{Calculation of the $L_\mu(\alpha)$ functions for large values of $\alpha$}
\label{subsec:lmularge}
Applications of the differential equation (\ref{lmudiff2}) are not limited to the small $\alpha$ expansion method 
presented
in the previous subsection. As the next offspring of Eq. (\ref{lmudiff2}) we shall present the large $\alpha$ asymptotic
expansion of the $L_\mu^0(\mu,\alpha)$ functions. As before, this expansion provides the starting values for the
recursive
relations (\ref{lgrowp}) and (\ref{lgrows}). To the best of our knowledge, no large $\alpha$ asymptotic expansion of
$L_\mu^0(\mu,\alpha)$ has been given in the previous works. It is because it is rather difficult to
derive this expansion having only the integral representation (\ref{lmuint}) at hand.

One may ask what is the point of deriving the large $\alpha$ asymptotic expansion of $L_\mu^0(\mu,\alpha)$ whereas the
analytical formula (\ref{lmugen}) is perfectly stable in this regime. Indeed, the larger the value of $\alpha$, the
more numerically stable Eq. (\ref{lmugen}) becomes. Therefore, it seems to be an unnecessary redundancy to introduce an
additional formula devoted specifically to the large $\alpha$ regime. This redundancy is only apparent, though. As one
shall see shortly, the final asymptotic formula is very simple. It can be implemented highly
efficiently and the calculation times are superior to the code based on Eq. (\ref{lmugen}). 
% Therefore, the asymptotic
% formula can be considered as a much better supplementation of the recursive method (subsection ...) than the explicit
% expression, Eq. (\ref{lmugen}).

In analogy with the previous subsection, the general solution of the differential equation (\ref{lmudiff2}) is given by
the
expression:
\begin{align}
\label{lmuginf}
D_\mu^{(1)}i_\mu(\alpha)+D_\mu^{(2)}k_\mu(\alpha)+\mathcal{L}^\infty_\mu(\alpha),
\end{align}
where $D_\mu^{(1)}$ and $D_\mu^{(2)}$ are new constants which we fix thereafter by using the initial conditions, and
$\mathcal{L}^\infty_\mu(\alpha)$ is a particular solution. Since we are now interested in the large $\alpha$ regime,
the latter takes the form
\begin{align}
\label{lmug2inf} 
\mathcal{L}^\infty_\mu(\alpha) = e^{-\alpha} \sum_{k=1}^\infty \frac{a_k^\mu}{\alpha^k}+
e^{-\alpha}\log(2\alpha)\sum_{k=1}^\infty \frac{b_k^\mu}{\alpha^k}.
\end{align}
Derivation of the recursion formulae obeyed by the coefficients $a_k^\mu$ and $b_k^\mu$ is done in the standard
fashion. The actual derivation is rather tedious and technical, so that we present only the final formulae:
\begin{align}
\label{recbinf}
b_{k+1}^\mu&=b_k^\mu\frac{\mu(\mu+1)-k(k-1)}{2k},\\
\label{recainf}
a_{k+1}^\mu&=\frac{b_k^\mu(2k-1)+2b_{k+1}^\mu-\left[ k(k-1)-\mu(\mu+1) \right]a_k^\mu}{2k},
\end{align}
for $k\geq1$, with an additional requirement $b_1^\mu=1/2$. The value of $a_1^\mu$ remains
arbitrary and we can make a conventional choice, analogous to the choice $c_\mu^\mu=0$ in the previous subsection.
Therefore, we put $a_1^\mu=0$ and then adjust properly the values of $D_\mu^{(1)}$ and
$D_\mu^{(2)}$, so that the initial conditions are automatically met. A closer look at Eq. (\ref{recbinf}) reveals
that $b_k^\mu$ with $k\geq\mu+2$ are zero. In other words, the the second summation in Eq. (\ref{lmug2inf}) actually
breaks off after $\mu+1$ terms. Despite that, the first summation remains infinite. The quenching of the
coefficients $b_k^\mu$ leads to simplifications in the formulae for $a_k^\mu$. One has
\begin{align}
\label{recainf2}
a_{\mu+2}^\mu=b_{\mu+1}^\mu\frac{2\mu+1}{2\mu+2},
\end{align}
and
\begin{align}
\label{recainf3}
a_{k+1}^\mu=a_k^\mu\frac{\mu(\mu+1)-k(k-1)}{2k},
\end{align}
for $k\geq\mu+2$. In practice, the coefficients $a_k^\mu$ and $b_k^\mu$ are optimally either tabulated or computed on 
the fly by using the following algorithm:
\begin{enumerate}
\item set $a_1^\mu=0$ and $b_1^\mu=1/2$,
\item build coefficients $b_k^\mu$ for $k\leq\mu+1$ by using Eq. (\ref{recbinf}),
\item build coefficients $a_k^\mu$ for $k\leq\mu+1$ by using Eq. (\ref{recainf}),
\item calculate $a_{\mu+2}^\mu$ from Eq. (\ref{recainf2}),
\item build the remaining coefficients $a_k^\mu$ up to a predefined value $k_{max}$ by using Eq. (\ref{recainf3}),
\item perform the final summations according to Eq. (\ref{lmug2inf}) and break off the first summation at $k_{max}$.
\end{enumerate}
The last issue is to determine the values of the constants $D_\mu^{(1)}$ and $D_\mu^{(2)}$. The reasoning for fixing
the first coefficient follows the same line as in the previous subsection. One sees from the integral representation
(\ref{lmuint}) that $L_\mu(\alpha)$ vanishes as $\alpha$ tends to infinity because the integrand in Eq. (\ref{lmuint})
dies off at the exponential rate. On the other hand, the function $i_\mu(\alpha)$ diverges as $\alpha$ tends to
infinity. Therefore, the constant $D_\mu^{(1)}$ must be equal to zero for every $\mu$. It is more difficult to fix the 
value of the second constant, $D_\mu^{(2)}$. As before, the corresponding derivation is included in Supplemental 
Material \cite{supplement}.

\begin{table*}[t]
\caption{Exemplary calculations of the $L_\mu^0(0,\alpha)$ functions for some representative values of $\alpha$. 
\emph{Exact}
denotes values calculated using Eq. (\ref{lmugen}) in extended arithmetic precision of 32 significant digits with the
\textsc{Mathematica} package (all digits shown are correct). \emph{Asymptotic expansion} column shows results of
calculations with Eqs. (\ref{lmuginf}) and (\ref{lmug2inf}) in the double precision arithmetic. \emph{Convergence}
denotes a number of terms in Eq. (\ref{lmug2inf}) required to converge the summation to 15 significant digits. The
symbol [k] denotes the powers of 10, 10$^k$.}
\begin{ruledtabular}
\begin{tabular}{cccc}
\label{table3}
$\mu$ & exact & asymptotic expansion & convergence \\
\hline\\[-2.1ex]
\multicolumn{4}{c}{ $\alpha=100.0$} \\
\hline\\[-2.1ex]
5  & 3.14 843 080 402 671 [$-46$] & 3.14 843 080 402 670 [$-46$] & 10 \\
10 & 1.61 980 042 035 663 [$-46$] & 1.61 980 042 035 663 [$-46$] & 12 \\
15 & 9.77 378 855 083 714 [$-47$] & 9.77 378 855 083 711 [$-47$] & 15 \\
20 & 6.46 965 882 608 025 [$-47$] & 6.46 965 882 608 030 [$-47$] & 19 \\
25 & 4.56 383 003 051 265 [$-47$] & 4.56 383 003 051 411 [$-47$] & 22 \\
30 & 3.37 452 175 547 398 [$-47$] & 3.37 452 175 547 026 [$-47$] & 25 \\
\hline\\[-2.1ex]
\multicolumn{4}{c}{ $\alpha=120.0$} \\
\hline\\[-2.1ex]
5  & 5.84 818 167 259 162 [$-55$] & 5.84 818 167 259 162 [$-55$] & 8 \\
10 & 3.09 500 781 737 830 [$-55$] & 3.09 500 781 737 831 [$-55$] & 10 \\
15 & 1.90 353 466 351 593 [$-55$] & 1.90 353 466 351 592 [$-55$] & 13 \\
20 & 1.27 691 236 663 178 [$-56$] & 1.27 691 236 663 178 [$-56$] & 16 \\
25 & 9.09 344 953 889 173 [$-56$] & 9.09 344 953 889 132 [$-56$] & 19 \\
30 & 6.77 029 861 923 809 [$-56$] & 6.77 029 861 923 817 [$-56$] & 22 \\[0.2ex]
\end{tabular}
\end{ruledtabular}
\end{table*}

% We have to to stress that a majority of the recursive formulae for the coefficients $a_k^\mu$ presented in this
% subsection is numerically unstable. These instabilities are not severe, though. Moreover, we do not require a large 
% number of the coefficients $a_k^\mu$ to be calculated since we are using the asymptotic method only for very large 
% values of $\alpha$. Under these circumstances, the infinite series in Eq. (\ref{lmug2inf}) converges rapidly. To verify 
% the validity of our expressions and to check the accuracy of the method we benchmarked it against ``exact'' values 
% calculated in an extended arithmetic precision using closed-form formulae. The results are presented in Table 
% \ref{table3}.

Let us now summarise the advances reported in the present section. The most important result is the differential
equation for the $L_\mu(\alpha)$ functions, Eq. (\ref{lmudiff2}). It has allowed us to derive both small and large 
$\alpha$ expansions of $L_\mu(\alpha)$. We proposed a practical
realisation of the recursive formula put forward by Harris. These three methods combined provide a new
way to calculate $L_\mu(\alpha)$ for all required values of the parameters. Finally, recurrence relations (\ref{lgrowp})
and (\ref{lgrows}) allow to build the final integrals $L_\mu^\sigma(p,\alpha)$ in a numerically stable fashion.
Therefore, we can conclude that the problem of accurate and robust calculation of
the $L_\mu^\sigma(p,\alpha)$ functions has been solved.

\section{Calculation of the $W_\mu^\sigma(p_1,p_2,\alpha_1,\alpha_2)$ functions}
\label{sec:wmu}
Before presenting our results, let us give a brief summary of the methods available in the literature for
the computation of the $W_\mu^\sigma(p_1,p_2,\alpha_1,\alpha_2)$ functions. In the early attempts, these integrals
resisted
to a direct integration and thus other schemes were proposed. Historically, the first fully analytical method was
published in the book of Kotani~\cite{kotani55} who established a family of simple recursion relations. Roughly
speaking, the major step of the Kotani recursions consists of building the integrals with larger values of $\mu$,
starting only with integrals with nonzero $p_1$, $p_2$, but $\mu=0$. The values of the
integrals $W_\mu^\sigma(p_1,p_2,\alpha_1,\alpha_2)$ grow very fast with the increasing $p_1$, $p_2$ but remain
approximately constant (or decrease) when the value of $\mu$ is enlarged.
Therefore, growing the value of $\mu$ at the cost of $p_1$ and $p_2$ is inherently connected with cancellation of huge
numbers to a relatively small result. This is the reason for a dramatic loss of digits observed when the
recursive process of Kotani is carried out up to large values of $\mu$. We can roughly estimate that for the present
purposes, Kotani scheme can only be used if the values of $\alpha_1$ and $\alpha_2$ are both very large, of
the order of 10-15. This is clearly highly unsatisfactory. 

As mentioned in the Introduction, Maslen and Trefry \cite{maslen90} derived analytical expressions for the
$W_\mu^\sigma(p_1,p_2,\alpha_1,\alpha_2)$ functions which is undoubtedly a large step forward. However,
these Authors failed to recognise some of the numerical instabilities connected with their expressions. The main working
formula of Maslen and Trefry, Eq. (\ref{wmugen}), cannot be used in practice for small values of $\alpha_2$. Examples
showing failure of this expression are presented further in the paper. Therefore, the
formulation of Maslen and Trefry alone cannot support large scale calculations, especially
when high angular momentum functions are present in the basis set. 

An alternative approach is based on a set of new recursive formulae proposed by Fern\'{a}ndez Rico \emph{et al}
\cite{rico94b}. In this scheme a set of auxiliary quantities is introduced and the so-called ``bisection'' algorithm is
used to carry out the recursion in two-dimensions, where the diagonal elements correspond to the
(scaled) $W_\mu^\sigma(p_1,p_2,\alpha_1,\alpha_2)$ functions. This recursive scheme is sufficient for small quantum
numbers, but becomes progressively less stable when quantum numbers are high, especially when the nonlinear coefficients
$\alpha_1$, $\alpha_2$ differ dramatically. Nonetheless, this scheme is elegant and straightforward, and has a 
potential of being robust which makes it suitable for small quantum numbers.

An important advance in the field is the 2002 work of Harris \cite{harris02}. Harris recognised the problems connected 
with the equations of Maslen and Trefry, and proposed new schemes for the computation of the 
$W_\mu^\sigma(p_1,p_2,\alpha_1,\alpha_2)$ functions. Small
$\alpha_2$ expansion of these integrals was considered but the working formula is not particularly useful,
mainly due to convergence problems and the necessity to calculate the $L_\mu^\sigma(p,\alpha)$ functions with very large
$p$. Another advance is the derivation of a new downward recursive scheme for 
the $W_\mu^\sigma(p_1,p_2,\alpha_1,\alpha_2)$ functions, completely disconnected from the method of Kotani. 
Unfortunately, the formulation of Harris is still not
free of problems. It has strong connections with the method of Fern\'{a}ndez Rico \emph{et al.} \cite{rico94b} and 
suffers from similar difficulties. Additionally, restarts in the downward recursion need to be carried out often.

Before passing further, let us recall an equation which appears in the recursive method of Kotani:
\begin{align}
\label{wmugrows}
\begin{split}
&W_\mu^{\sigma+1}(p_1,p_2,\alpha_1,\alpha_2)=
\frac{(\mu-\sigma)(\mu-\sigma+1)^2}{2\mu+1}\\
&\times W_{\mu+1}^\sigma(p_1,p_2,\alpha_1,\alpha_2)
-(\mu-\sigma)(\mu+\sigma+1)\\
&\times W_\mu^\sigma(p_1+1,p_2+1,\alpha_1,\alpha_2)
+\frac{(\mu+\sigma+1)(\mu+\sigma)^2}{2\mu+1}\\
&\times W_{\mu-1}^\sigma(p_1,p_2,\alpha_1,\alpha_2).
\end{split}
\end{align}
The above expression is sufficiently numerically stable for all relevant values of the parameters. Therefore, it 
provides an efficient and reliable method of generation of $W_\mu^\sigma(p_1,p_2,\alpha_1,\alpha_2)$ starting only with 
the integrals with $\sigma=0$. Further in the article, we are concerned only with calculation 
of $W_\mu^0(p_1,p_2,\alpha_1,\alpha_2)$.

\subsection{The differential equation for the $W_\mu^0(p_1,p_2,\alpha_1,\alpha_2)$ functions}
\label{subsec:wmudiff}
In this subsection we shall derive a differential equation obeyed by the $W_\mu^0(p_1,p_2,\alpha_1,\alpha_2)$
functions. Derivations follow roughly the same idea as the one given in Subsection \ref{subsec:lmusmall}, but are 
considerably more complicated. Fortunately, it also leads to an unexpectedly simple result. We introduce the 
abbreviation $W_\mu(p_1,p_2,\alpha_1,\alpha_2):=W_\mu^0(p_1,p_2,\alpha_1,\alpha_2)$.

Let us begin with the well-know differential equation for the Legendre functions:
\begin{align}
(1-\xi_2^2)P_\mu''(\xi_2)-2\xi_2P_\mu'(\xi_2)+\mu(\mu+1)P_\mu(\xi_2)=0.
\end{align}
By multiplying by $e^{-\alpha_2\xi_2}$ and integrating over the interval [1,$\,\xi_1$] one arrives at:
\begin{align}
\begin{split}
0&=\int_1^{\xi_1} P_\mu''(\xi_2)e^{-\alpha_2\xi_2}-\int_1^{\xi_1} \xi_2^2 P_\mu''(\xi_2)e^{-\alpha_2\xi_2}\\
&-2\int_1^{\xi_1} \xi_2 P_\mu'(\xi_2)e^{-\alpha_2\xi_2}+\mu(\mu+1)\int_1^{\xi_1} P_\mu(\xi_2)e^{-\alpha_2\xi_2}.
\end{split}
\end{align}
The first and the second of the integrals need now to be integrated by parts twice. Noting that $P_\mu(\xi_2)$ is
regular at $\xi_2=1$ one finds
\begin{align}
\begin{split}
&\int_1^{\xi_1} P_\mu''(\xi_2)e^{-\alpha_2\xi_2}=
e^{-\alpha_2\xi_1}\Big[P_\mu'(\xi_1) +\alpha_2 P_\mu(\xi_1)\Big]\\
&-e^{-\alpha_2}\Big[P_\mu'(1)+\alpha_2P_\mu(1)\Big]
+\alpha_2^2\int_1^{\xi_1}P_\mu(\xi_2)e^{-\alpha_2\xi_2},
\end{split}
\end{align}
and
\begin{align}
\begin{split}
&\int_1^{\xi_1} \xi_2^2 P_\mu''(\xi_2)e^{-\alpha_2\xi_2}=-2\int_1^{\xi_1} \xi_2
P_\mu'(\xi_2)e^{-\alpha_2\xi_2}\\
&\xi_1^2e^{-\alpha_2\xi_1}\Big[P_\mu'(\xi_1)+\alpha_2P_\mu(\xi_1)\Big]
-e^{-\alpha_2}\Big[P_\mu'(1)+\alpha_2P_\mu(1)\Big]\\
&-2\alpha_2\int_1^{\xi_1} \xi_2 P_\mu(\xi_2)e^{-\alpha_2\xi_2}
+\alpha_2^2\int_1^{\xi_1} \xi_2^2 P_\mu(\xi_2)e^{-\alpha_2\xi_2}.
\end{split}
\end{align}
By inserting these identities into the initial equation many cancellations occur and finally we obtain
\begin{align}
\begin{split}
&(1-\xi_1^2)P_\mu'(\xi_1)e^{-\alpha_2\xi_1}+\alpha_2 (1-\xi_2^2)P_\mu(\xi_1)e^{-\alpha_2\xi_1}\\
&+2\alpha_2\int_1^{\xi_1} \xi_2 P_\mu(\xi_2)e^{-\alpha_2\xi_2}
-\alpha_2^2\int_1^{\xi_1} \xi_2^2 P_\mu(\xi_2)e^{-\alpha_2\xi_2}\\
&+\Big[\mu(\mu+1)+\alpha_2^2\Big]\int_1^{\xi_1} P_\mu(\xi_2)e^{-\alpha_2\xi_2}=0.
\end{split}
\end{align}
The next step is to multiply both sides of the above equation by $Q_\mu(\xi_1)\xi_1^{p_1}e^{-\alpha_1\xi_1}$ and
integrate by $\xi_1$ over the interval [$1,+\infty$). Additionally, we make use of the identity
\begin{align}
\label{derivp}
(1-\xi_1^2)P_\mu'(\xi_1)=\mu P_{\mu-1}(\xi_1)-\mu\xi_1P_\mu(\xi_1),
\end{align}
to arrive at
\begin{align}
\label{wmudiff1}
\begin{split}
&\alpha_2^2\,\frac{\partial^2}{\partial\alpha_2^2}w_\mu(p_1,0,\alpha_1,\alpha_2)+2\alpha_2\,\frac{\partial}{
\partial\alpha_2}w_\mu(p_1,0,\alpha_1,\alpha_2)\\
&-\Big[\mu(\mu+1)+\alpha_2^2\Big]w_\mu(p_1,0,\alpha_1,\alpha_2)=\\
&+\mu\, T_{\mu-1,\mu}(p_1,\alpha_1+\alpha_2)-\mu\, T_{\mu\mu}(p_1+1,\alpha_1+\alpha_2)\\
&+\alpha_2 T_{\mu\mu}(p_1,\alpha_1+\alpha_2)-\alpha_2 T_{\mu\mu}(p_1+2,\alpha_1+\alpha_2).
\end{split}
\end{align}
where we have introduced, in analogy to Harris \cite{harris02}, a new function family $T_{\mu\nu}(p,\alpha)$
\begin{align}
T_{\mu\nu}(p,\alpha)=\int_1^\infty d\xi P_\mu(\xi)Q_\nu(\xi)\xi^p e^{-\alpha \xi}.
\end{align}
The above equation is a differential equation for $w_\mu(p_1,0,\alpha_1,\alpha_2)$ with respect to
$\alpha_2$. It has probably been unknown thus far. The remaining effort is to obtain the necessary series expansion 
valid in the small $\alpha_2$ regime. Unfortunately, the
above differential equation is not well suited for further developments because of the complicated form of the
inhomogeneous term on the right hand side. The resulting small $\alpha_2$ expansion is complicated, slowly convergent
and expensive to calculate. Additionally, the necessity to compute the $T_{\mu\nu}(p,\alpha)$ functions is a 
disadvantage. Therefore, we must seek a reformulation of some kind which allows a more convenient
numerical evaluation.                                                                                                   
           
The second part of the derivation starts with the observation that $w_\mu(0,p_1,\alpha_2,\alpha_1)$ can be
cast in the following equivalent form
\begin{align}
\label{weq}
\begin{split}
w_\mu(0,p_1,\alpha_2,\alpha_1)&=\int_1^\infty d\xi_2 P_\mu(\xi_2)\xi_2^{p_1}e^{-\alpha_1\xi_2}\\
&\times\int_{\xi_2}^\infty Q_\mu(\xi_1)e^{-\alpha_2\xi_1}.
\end{split}
\end{align}
Starting again with the differential equation (\ref{eqq}), we multiply both sides by $e^{-\alpha_2\xi_1}$ and integrate
over $\xi_1$ on the interval [$\xi_2,+\infty$). The next step of the derivation is exactly the same as previously, the
first two integrals are integrated by parts twice. The resulting expressions are inserted back into the initial
equation. This procedure was described in detail earlier so here we list only the result:
\begin{align}
\begin{split}
&-(1-\xi_2^2)Q_\mu'(\xi_2)e^{-\alpha_2\xi_2}-\alpha_2(1-\xi_2^2)Q_\mu(\xi_2)e^{-\alpha_2\xi_2}\\
&+\alpha_2^2\int_{\xi_2}^\infty Q_\mu(\xi_1)e^{-\alpha_2\xi_1}+2\alpha\int_{\xi_2}^\infty
Q_\mu(\xi_1)\xi_1e^{-\alpha_2\xi_1}\\
&-\alpha_2^2\int_{\xi_2}^\infty Q_\mu(\xi_1)\xi_1^2e^{-\alpha_2\xi_1}=0.
\end{split}
\end{align}
Noting that exactly the same expression as (\ref{derivp}) holds also for the Legendre functions of the second kind,
$Q_\mu$, we can get rid of the derivative in the first term of the above expression. Next, we multiply both sides by
$P_\mu(\xi_2)\xi_2^{p_1}e^{-\alpha_1\xi_2}$ and integrate over $\xi_2$ on the interval [$1,+\infty$). By invoking Eq.
(\ref{weq}) one can bring the final result into the form
\begin{align}
\label{wmudiff2}
\begin{split}
&\alpha_2^2\,\frac{\partial^2}{\partial\alpha_2^2}w_\mu(0,p_1,\alpha_2,\alpha_1)+2\alpha_2\,\frac{\partial}{
\partial\alpha_2}w_\mu(0,p_1,\alpha_2,\alpha_1)\\
&-\Big[\mu(\mu+1)+\alpha_2^2\Big]w_\mu(0,p_1,\alpha_2,\alpha_1)=\\
&-\mu\, T_{\mu,\mu-1}(p_1,\alpha_1+\alpha_2)+\mu\, T_{\mu\mu}(p_1+1,\alpha_1+\alpha_2)\\
&-\alpha_2 T_{\mu\mu}(p_1,\alpha_1+\alpha_2)+\alpha_2 T_{\mu\mu}(p_1+2,\alpha_1+\alpha_2),
\end{split}
\end{align}
which constitutes the second required ingredient. Now, Eqs. (\ref{wmudiff1}) and (\ref{wmudiff2}) are added together,
and by making use of Eq. (\ref{bigw}) one finds
\begin{align}
\begin{split}
&\alpha_2^2\,\frac{\partial^2}{\partial\alpha_2^2}W_\mu(p_1,0,\alpha_1,\alpha_2)+2\alpha_2\,\frac{\partial}{
\partial\alpha_2}W_\mu(p_1,0,\alpha_1,\alpha_2)\\
&-\Big[\mu(\mu+1)+\alpha_2^2\Big]W_\mu(p_1,0,\alpha_1,\alpha_2)=\\
&\mu\, T_{\mu-1,\mu}(p_1,\alpha_1+\alpha_2)-\mu\, T_{\mu,\mu-1}(p_1,\alpha_1+\alpha_2).
\end{split}
\end{align}
To finally get rid of the $T_{\mu\nu}$ functions in the inhomogeneous term let us recall the transfer relation between
the Legendre functions:
\begin{align}
P_{\mu+1}(\xi)Q_\mu(\xi)-P_\mu(\xi)Q_{\mu+1}(\xi)=\frac{\Gamma(\mu+1)}{\Gamma(\mu+2)},
\end{align}
which gives
\begin{align}
\begin{split}
&T_{\mu-1,\mu}(p_1,\alpha_1+\alpha_2) - T_{\mu,\mu-1}(p_1,\alpha_1+\alpha_2)=\\
&-\frac{1}{\mu}A_{p_1}(\alpha_1+\alpha_2).
\end{split}
\end{align}
By inserting this expression into the differential equation (\ref{wmudiff2}) one obtains the most important
formula of this work
%\begin{widetext}
\begin{align}
\label{bigwdiff}
\begin{split}
&\alpha_2^2\,\frac{\partial^2}{\partial\alpha_2^2}W_\mu(p_1,0,\alpha_1,\alpha_2)+2\alpha_2\,\frac{\partial}{
\partial\alpha_2}W_\mu(p_1,0,\alpha_1,\alpha_2)\\
&-\Big[\mu(\mu+1)+\alpha_2^2\Big]W_\mu(p_1,0,\alpha_1,\alpha_2)=-A_{p_1}(\alpha_1+\alpha_2).
\end{split}
\end{align} 
%\end{widetext}
Noting that $W_\mu(p_1,0,\alpha_1,\alpha_2)$ is a complicated function of many variables, the simplicity of Eq.
(\ref{bigwdiff}) is somehow surprising. First of all, we already know the solution of the homogeneous equation and it
is the same as for the $L_\mu(\alpha)$ functions. The inhomogeneous term on the right hand side is also a simple
function which has a potential of providing reasonably uncomplicated expansions. In the next subsection we deal with the
small $\alpha_2$ expansion of $W_\mu(p_1,0,\alpha_1,\alpha_2)$, starting with the differential equation
(\ref{bigwdiff}).

\subsection{Calculation of the $W_\mu^0(p_1,p_2,\alpha_1,\alpha_2)$ functions for small values of $\alpha_2$}
\label{subsec:wmusmall}
Similarly as in Section \ref{subsec:lmusmall}, the solution of the differential equation (\ref{bigwdiff}) can be 
written in the form
\begin{align}
\label{partw2}
C_{\mu
p_1}^{(1)}(\alpha_1)i_\mu(\alpha_2)+C_{\mu p_1}^{(2)}(\alpha_1)k_\mu(\alpha_2)+\mathcal{W}_\mu^{p_1}(\alpha_1,\alpha_2),
\end{align}
but the constants $C_{\mu p_1}^{(1)}(\alpha_1)$ and $C_{\mu p_1}^{(2)}(\alpha_1)$ are now dependent on the value of
$\alpha_1$ \emph{i.e.} they are no longer discrete quantities. This leads to huge complications during their evaluation
which shall be considered further. The function $\mathcal{W}_\mu^{p_1}(\alpha_1,\alpha_2)$ can be assumed to have
the following series expansion in the small $\alpha_2$ regime
\begin{align}
\label{partw}
\mathcal{W}_\mu^{p_1}(\alpha_1,\alpha_2)= \sum_{k=0}^\infty c_k^{\mu p_1} \alpha_2^k +
\log(2\alpha_2)\sum_{k=\mu(+2)}^\infty d_k^{\mu p_1} \alpha_2^k.
\end{align}
Formally, one should set $c_k^{\mu p_1}:=c_k^{\mu p_1}(\alpha_1)$ since the expansion coefficients are functions of
$\alpha_1$. However, this dependence is obvious and we decided to suppress it in order to make our
equations more compact. The inhomogeneous term in Eq. (\ref{bigwdiff}) possesses the small $\alpha_2$
expansion
\begin{align}
\label{rhsserie}
-A_{p_1}(\alpha_1+\alpha_2) = \sum_{k=0}^\infty (-1)^{k+1} A_{p_1+k}(\alpha_1)\alpha_2^k.
\end{align}
To find the recursion relations for the coefficients $c_k^{\mu p_1}$ and $d_k^{\mu p_1}$ one has to insert the formula
(\ref{partw}) into Eq. (\ref{bigwdiff}) and proceed in exactly the same way as in Subsection \ref{subsec:lmusmall}. In 
fact, the only
difference between these derivations lies in a small difference between the expansions of the inhomogeneous terms. 
Therefore,
there is no point in repeating this derivation here and we confine ourselves to the presentation of the final results.
One first builds $c_k^{\mu p_1}$ coefficients up to, and including, $k=\mu-1$ by using the formula
\begin{align}
c_k^{\mu p_1}\Big[k(k+1)-\mu(\mu+1)\Big]-c_{k-2}^{\mu p_1}+(-1)^k A_{p_1+k}(\alpha_1)=0,
\end{align}
with the initial values being $c_0^{\mu p_1}=A_{p_1}(\alpha_1)/\mu(\mu+1)$ and $c_1^{\mu
p_1}=-\frac{-A_{p_1+1}(\alpha_1)}{2-\mu(\mu+1)}$. The first of the coefficients in the logarithmic part of the expansion
is found from the relation
\begin{align}
d_\mu^{\mu p_1}(2\mu+1)-c_{\mu-2}^{\mu p_1}+(-1)^\mu A_{p_1+\mu}(\alpha_1)=0,
\end{align}
and then the other coefficients are calculated recursively as
\begin{align}
d_k^{\mu p_1}=\frac{d_{k-2}^{\mu p_1}}{k(k+1)-\mu(\mu+1)}.
\end{align}
Finally, $c_k^{\mu p_1}$ coefficients with $k\geq \mu+1$ are build from the formula
\begin{align}
\begin{split}
&c_k^{\mu p_1}\Big[k(k+1)-\mu(\mu+1)\Big]+d_k^{\mu p_1}(2k+1)\\
&-c_{k-2}^{\mu p_1}+(-1)^k A_{p_1+k}(\alpha_1)=0.
\end{split}
\end{align}
The choice of $c_\mu^{\mu p_1}$ is conventional and we can safely put it equal to zero, as discussed earlier.
The remaining problem is the determination of the constants $C_{\mu p_1}^{(1)}(\alpha_1)$ and $C_{\mu 
p_1}^{(2)}(\alpha_1)$. Using a similar reasoning as utilised previously, the value of $C_{\mu p_1}^{(2)}(\alpha_1)$ can 
immediately be fixed at zero. However, the derivation of an analytical formula for $C_{\mu p_1}^{(1)}(\alpha_1)$ is 
much more cumbersome and is presented in Supplemental Material \cite{supplement}.

The small parameter expansions given here and in Section \ref{subsec:lmusmall} seem to be completely analogous since 
the working formulae
differ only by the presence of the $A_k$ functions. There is, however, a big difference that practically limits the
usefulness of the formula (\ref{partw}). The inhomogeneous term in the differential equation (\ref{lmudiff2}) has a
small $\alpha$ Taylor expansion which is convergent for all relevant values of the parameter. Conversely, the series 
on
the right hand side of Eq. (\ref{rhsserie}) has a finite radius of convergence. Namely, it is convergent if and only if
the inequality $\alpha_2<\alpha_1$ holds. From the mathematical point of view, when $\alpha_1<\alpha_2$ one can make
use of the symmetry relation
\begin{align}
\label{wmusymm}
W_\mu(p_1,p_2,\alpha_1,\alpha_2)=W_\mu(p_2,p_1,\alpha_2,\alpha_1),
\end{align}
so that the roles of $\alpha_1$ and $\alpha_2$ are exchanged and the resulting series (\ref{rhsserie}) falls within the
convergence region. Unfortunately, the practical situation is more complex. It is understandable that when
$\alpha_2$ becomes close to $\alpha_1$ the series (\ref{rhsserie}) converges progressively slower. As a result, the
series in Eq. (\ref{partw}) also suffers from the pathologically slow convergence pattern. This makes the presented
method virtually useless unless $\alpha_1$ and $\alpha_2$ are reasonably spaced. Our numerical experience shows that
the difference $|\alpha_1-\alpha_2|$ must be larger than $2$ to ensure a sufficiently fast rate of convergence.

\begin{table*}
\caption{Exemplary calculations of the $W_\mu(p_1,p_2,\alpha_1,\alpha_2)$ functions for a few representative values of
$\alpha_1$ and $\alpha_2$. \emph{Exact} denotes values calculated using Eq. (\ref{wmugen}) in the extended arithmetic
precision of 120 significant digits with the \textsc{Mathematica} package (all digits shown are correct).
\emph{Closed-form} denotes calculations with Eq. (\ref{wmugen}) in the double precision arithmetic (around 15
significant figures). \emph{New} column shows results of calculations with Eqs. (\ref{partw}) and (\ref{partw2}), also
in the double
precision arithmetic. \emph{Convergence} denotes a number of terms in Eq. (\ref{partw}) required to converge both
summations to 15 significant digits. The symbol $[k]$ denotes the powers of 10, 10$^k$.}
\begin{ruledtabular}
\begin{tabular}{ccccc}
\label{table4}
$\mu$ & exact & closed-form & new & convergence \\
\hline\\[-2.1ex]
\multicolumn{5}{c}{ $\alpha_1=3.0$, $\alpha_2=0.5$, $p_1=0$, $p_2=0$} \\
\hline\\[-2.1ex]
0  & 1.04 486 860 277 951 [$-$02] & 1.04 486 860 277 951 [$-$02] & 1.04 486 860 277 951 [$-$02] & 18\\
5  & 2.77 344 623 535 900 [$-$04] & 2.77 344 591 894 414 [$-$04] & 2.77 344 623 535 900 [$-$04] & 20\\
10 & 7.76 549 171 325 524 [$-$05] & 2.30 066 585 106 053 [$+$00] & 7.76 549 171 325 524 [$-$05] & 21\\
15 & 3.57 847 552 224 820 [$-$05] & 1.28 381 137 541 490 [$+$12] & 3.57 847 552 224 820 [$-$05] & 22\\
20 & 2.04 886 403 945 215 [$-$05] & 1.65 546 870 529 827 [$+$28] & 2.04 886 403 945 215 [$-$05] & 22\\
25 & 1.32 510 984 698 693 [$-$05] & 1.05 032 000 000 000 [$+$44] & 1.32 510 984 698 693 [$-$05] & 23\\
\hline\\[-2.1ex]
\multicolumn{5}{c}{ $\alpha_1=10.0$, $\alpha_2=2.0$, $p_1=0$, $p_2=0$} \\
\hline\\[-2.1ex]
0  & 3.06 472 238 344 757 [$-$07] & 3.06 472 238 344 757 [$-$07] & 3.06 472 238 344 757 [$-$07] & 24\\
5  & 1.53 355 187 887 866 [$-$08] & 1.53 355 187 883 772 [$-$08] & 1.53 355 187 887 865 [$-$08] & 27\\
10 & 4.50 949 894 593 816 [$-$09] & 4.50 918 163 373 806 [$-$09] & 4.50 949 894 593 816 [$-$09] & 29\\
15 & 2.10 206 354 777 336 [$-$09] & 2.30 624 503 190 029 [$-$05] & 2.10 206 354 777 336 [$-$09] & 29\\
20 & 1.20 875 103 359 696 [$-$09] & 1.80 309 371 523 750 [$+$04] & 1.20 875 103 359 697 [$-$09] & 30\\
25 & 7.83 382 082 527 984 [$-$10] & 1.00 728 429 616 952 [$+$14] & 7.83 382 082 527 983 [$-$10] & 30\\
\hline\\[-2.1ex]
\multicolumn{5}{c}{ $\alpha_1=3.0$, $\alpha_2=0.5$, $p_1=5$, $p_2=0$} \\
\hline\\[-2.1ex]
0  &  7.48 701 970 608 968 [$-$02] & 7.48 701 970 608 967 [$-$02] & 7.48 701 970 608 968 [$-$02] & 23\\
5  &  1.79 908 205 094 134 [$-$03] & 1.79 908 423 520 203 [$-$03] & 1.79 908 205 094 134 [$-$03] & 26\\
10 &  5.03 737 212 031 091 [$-$04] & 7.00 669 096 089 900 [$+$02] & 5.03 737 212 031 091 [$-$04] & 27\\
15 &  2.32 162 471 967 834 [$-$04] & 5.29 745 051 970 872 [$+$15] & 2.32 162 471 967 834 [$-$04] & 27\\
20 &  1.32 933 530 186 838 [$-$04] & 3.87 308 518 932 093 [$+$31] & 1.32 933 530 186 838 [$-$04] & 28\\
25 &  8.59 780 135 199 690 [$-$05] & 1.22 261 970 344 498 [$+$47] & 8.59 780 135 199 689 [$-$05] & 28\\
\hline\\[-2.1ex]
\multicolumn{5}{c}{ $\alpha_1=10.0$, $\alpha_2=2.0$, $p_1=5$, $p_2=0$} \\
\hline\\[-2.1ex]
0  & 5.18 010 434 002 219 [$-$07] & 5.18 010 434 002 219 [$-$07] & 5.18 010 434 002 218 [$-$07] & 27\\
5  & 2.46 474 533 411 348 [$-$08] & 2.46 474 533 415 337 [$-$08] & 2.46 474 533 411 347 [$-$08] & 30\\
10 & 7.21 451 958 797 078 [$-$09] & 7.21 568 303 718 723 [$-$09] & 7.21 451 958 797 074 [$-$09] & 32\\
15 & 3.35 931 293 998 890 [$-$09] & 1.15 756 754 553 331 [$-$04] & 3.35 931 293 998 888 [$-$09] & 33\\
20 & 1.93 091 293 856 568 [$-$09] & 1.52 583 622 236 550 [$+$05] & 1.93 091 293 856 568 [$-$09] & 33\\
25 & 1.25 116 286 073 580 [$-$09] & 1.49 697 488 157 192 [$+$16] & 1.25 116 286 073 579 [$-$09] & 34\\
\hline\\[-2.1ex]
\multicolumn{5}{c}{ $\alpha_1=3.0$, $\alpha_2=0.5$, $p_1=0$, $p_2=5$} \\
\hline\\[-2.1ex]
0  & 1.28 329 165 081 863 [$+$01] & 1.28 329 165 081 863 [$+$01] & 1.28 329 165 081 863 [$+$01] & 29\\
5  & 3.31 860 127 430 244 [$-$03] & 3.30 708 670 298 918 [$-$03] & 3.31 860 127 430 243 [$-$03] & 29\\
10 & 5.87 022 662 870 030 [$-$04] & 1.06 012 946 964 569 [$+$07] & 5.87 022 662 870 029 [$-$04] & 31\\
15 & 2.48 544 920 341 368 [$-$04] & 5.63 197 344 090 803 [$+$20] & 2.48 544 920 341 368 [$-$04] & 32\\
20 & 1.38 157 704 830 518 [$-$04] & 1.20 667 417 809 212 [$+$35] & 1.38 157 704 830 518 [$-$04] & 32\\
25 & 8.81 350 672 621 061 [$-$04] & 6.40 686 820 917 187 [$+$51] & 8.81 350 672 621 062 [$-$05] & 33\\
\hline\\[-2.1ex]
\multicolumn{5}{c}{ $\alpha_1=10.0$, $\alpha_2=2.0$, $p_1=0$, $p_2=5$} \\
\hline\\[-2.1ex]
0  & 3.58 469 358 658 655 [$-$06] & 3.58 469 358 658 655 [$-$06] & 3.58 469 358 658 655 [$-$06] & 33\\
5  & 3.25 647 646 961 604 [$-$08] & 3.25 647 645 628 771 [$-$08] & 3.25 647 646 961 624 [$-$08] & 33\\
10 & 7.83 161 224 394 686 [$-$09] & 1.21 883 116 932 509 [$-$08] & 7.83 161 224 394 619 [$-$09] & 35\\
15 & 3.48 945 279 149 928 [$-$09] & 6.77 701 483 960 846 [$-$01] & 3.48 945 279 149 931 [$-$09] & 38\\
20 & 1.97 344 359 497 490 [$-$09] & 9.75 387 030 280 981 [$+$08] & 1.97 344 359 497 492 [$-$09] & 38\\
25 & 1.26 892 528 802 273 [$-$09] & 2.09 712 604 451 529 [$+$19] & 1.26 892 528 802 274 [$-$09] & 39\\
\hline\\[-2.1ex]
\multicolumn{5}{c}{ $\alpha_1=3.0$, $\alpha_2=0.5$, $p_1=5$, $p_2=5$} \\
\hline\\[-2.1ex]
0  & 1.16 382 213 456 748 [$+$02] & 1.16 382 213 456 748 [$+$02] & 1.16 382 213 456 748 [$+$02] & 35\\
5  & 1.85 882 259 866 799 [$-$01] & 1.87 725 859 472 266 [$-$01] & 1.85 882 259 866 784 [$-$01] & 33\\
10 & 3.82 726 511 424 708 [$-$02] & 9.29 543 672 424 714 [$+$09] & 3.82 726 511 424 750 [$-$02] & 35\\
15 & 1.64 898 677 239 583 [$-$02] & 9.02 398 949 011 648 [$+$23] & 1.64 898 677 239 591 [$-$02] & 37\\
20 & 9.21 265 882 301 559 [$-$03] & 7.84 436 191 383 490 [$+$39] & 9.21 265 882 301 558 [$-$03] & 37\\
25 & 5.88 975 925 491 820 [$-$03] & 2.12 258 315 835 870 [$+$56] & 5.88 975 925 491 821 [$-$03] & 38\\
\hline\\[-2.1ex]
\multicolumn{5}{c}{ $\alpha_1=10.0$, $\alpha_2=2.0$, $p_1=5$, $p_2=5$} \\
\hline\\[-2.1ex]
0  & 6.31 318 894 312 804 [$-$06] & 6.31 318 894 312 804 [$-$06] & 6.31 318 894 312 804 [$-$06] & 37\\
5  & 6.58 645 047 384 616 [$-$08] & 6.58 645 051 424 336 [$-$08] & 6.58 645 047 384 616 [$-$08] & 37\\
10 & 1.61 556 791 342 107 [$-$08] & 3.12 085 735 032 497 [$-$08] & 1.61 556 791 341 897 [$-$08] & 38\\
15 & 7.22 199 212 246 876 [$-$09] & 1.23 801 100 582 205 [$+$01] & 7.22 199 212 246 509 [$-$09] & 41\\
20 & 4.08 869 377 197 726 [$-$09] & 3.81 270 930 437 817 [$+$10] & 4.08 869 377 197 785 [$-$09] & 43\\
25 & 2.63 027 491 413 946 [$-$09] & 1.01 348 114 811 063 [$+$22] & 2.63 027 491 413 945 [$-$09] & 43\\
\end{tabular}
\end{ruledtabular}
\end{table*}

Despite this shortcoming, the presented method solves a large majority of the problems connected with the small
$\alpha_2$ regime. Let us account for this statement by using the simplest possible example. For typical basis sets and
reasonable values of the internuclear distances, only a handful of functions in the basis set can give rise to the
values of $\alpha$ which fall in the problematic regime. Therefore, the number of integrals in which both $\alpha_1$ and
$\alpha_2$ are small constitutes only a few percent, or even less, of the total number of integrals to be
evaluated. On the other hand, the number of possible combinations in which $\alpha_2$ is small but $\alpha_1$ is large
or moderate (or \emph{vice versa}) is at least an order of magnitude larger. Typically, this situation corresponds to
10-20\% of the total number of integrals which is definitely a significant fraction. The latter combination of
$\alpha_2$ and $\alpha_1$ is perfectly suited for the present algorithm since in most cases the difference
$|\alpha_1-\alpha_2|$ is sufficiently large. Of course, the larger this difference is, the faster the series in Eq.
(\ref{partw}) converge.

A slight inconvenience connected with Eq. (\ref{partw}) is that it includes explicitly only integrals with $p_2=0$.
Higher values of $p_2$ have to be calculated by a consecutive differentiation with respect to $\alpha_2$. Series present
in Eq. (\ref{partw}) are trivial to differentiate analytically but the resulting series converge slightly slower.
However, since the expansion coefficients $c_k^{\mu p_1}$ and $d_k^{\mu p_1}$ are shared between the integrals with
different values of $p_2$ they have to be calculated only once. Therefore, the integrals with higher values of $p_2$ 
can be
calculated at a small additional cost, once a sufficiently large number of the expansion coefficients has been
calculated in advance.

In Table \ref{table4} we present the calculated values of $W_\mu(p_1,p_2,\alpha_1,\alpha_2)$ for a selected set of
$\alpha_1$ and $\alpha_2$. We included two the most challenging cases: when the difference between $\alpha_1$ and
$\alpha_2$ is small, and when this difference is larger but also the value of $\alpha_2$ is larger. In both cases one
could expect problems with convergence of the expansion or a loss of digits during the calculations. However, it turns
out that for a reasonably wide range of $\mu$, $p_1$ and $p_2$, our method provides an accuracy of at least 12-13
digits, and even more on the average. The number of terms needed to converge both summations in Eq. (\ref{partw}) is of
order of few tens. This is acceptable, taking into consideration that the coefficients of the expansion are calculated
efficiently by a fast and stable recursive process. Calculation of the constant $C_\mu^{(1)}(\alpha_1)$ (see 
Supplemental Material \cite{supplement}
for the accompanying discussion) consumes a significant fraction of the computational time. However, if a fast routine
for the calculation of $L_\mu(p,\alpha)$ is provided, the overhead is still acceptable. To sum up, the series expansion
method is superior to the analytical scheme which basically breaks down once the borderline of $\mu=5-10$ has been
crossed. 

\subsection{Calculation of the $W_\mu^0(p_1,p_2,\alpha_1,\alpha_2)$ functions for large values of $\alpha_2$}
\label{subsec:wmularge}
The remaining formula which can straightforwardly be derived from the differential equation (\ref{bigwdiff}) is the
asymptotic expansion of $W_\mu^0(p_1,p_2,\alpha_1,\alpha_2)$ for large values of $\alpha_2$. This method is designed
mainly to reduce costs of the calculations since the analytical expression, Eq. (\ref{wmugen}), is stable in this
regime. However, as the values of $\alpha_i$ become large, one can expect integrals with comparable values of
$\beta_i$. As mentioned earlier, in such cases the ellipsoidal expansion converges slower and quite large values of
$\mu$ are required to achieve a desired accuracy. In this light, any method that significantly reduces the costs of
the calculations for large $\alpha_i$ is definitely welcomed.

To start the derivation, we first require an expression that defines the asymptotic behaviour of the inhomogeneity in
Eq. (\ref{bigwdiff}). It has the following form:
\begin{align}
\label{aplarge1}
A_{p_1}(\alpha_1+\alpha_2)=e^{-\alpha_1-\alpha_2}\sum_{k=0}^\infty \frac{(-1)^k}{\alpha_2^{k+1}}
\mathcal{C}_{kp_1}(\alpha_1),
\end{align}
where the coefficients $\mathcal{C}_{kp_1}(\alpha_1)$ are simple polynomials in $\alpha_1$:
\begin{align}
\label{aplarge2}
\mathcal{C}_{kp_1}(\alpha_1)=\sum_{l=0}^{\mbox{\small min}(k,p_1)} \frac{p_1}{(p_1-l)!} (-1)^l \alpha_1^{k-l} { k
\choose l }.
\end{align}
We did not manage to further simplify the above expression. However, it is clear that the coefficients
$\mathcal{C}_{kp_1}(\alpha_1)$ are independent of $\mu$ and $p_2$ and therefore they need to be calculated only once for
a given set of the $W_\mu^0(p_1,p_2,\alpha_1,\alpha_2)$ integrals.
% Eqs. (\ref{aplarge1}) and (\ref{aplarge2}) can be derived by first noting that $A_n(x)=E_{-n}(x)$, where $E_n$ is the 
% generalised exponential integral function. Then, by using the well-known asymptotic expansion of $E_n$, the final 
% formula is readily achieved.

Our Ansatz for the large $\alpha_2$ asymptotic solution of the differential equation (\ref{bigwdiff}) is as follows
\begin{align}
D_\mu^{(1)}(\alpha_1)i_\mu(\alpha_2)+D_\mu^{(2)}(\alpha_1)k_\mu(\alpha_2)+\mathcal{W}_\mu^{p_1,\infty}(\alpha_1,
\alpha_2),
\end{align}
where the particular solution $\mathcal{W}_\mu^{p_1,\infty}$ is given by the inverse power expansion in $\alpha_2$
multiplied by the proper exponential term:
\begin{align}
\label{wmuexp}
\mathcal{W}_\mu^{p_1,\infty}(\alpha_1,\alpha_2) = e^{-\alpha_1-\alpha_2} \sum_{k=0}^\infty
\frac{a_k^{\mu p_1}}{\alpha_2^{k+1}},
\end{align}
In the above expression, $a_k^{\mu p_1}$ are implicitly assumed to be functions of $\alpha_1$ and the corresponding
notation was
suppressed for brevity. The necessary recursive relation for $a_k^{\mu p_1}$ is found by inserting the above formula
into the
differential equation (\ref{bigwdiff}) and grouping the same inverse powers of $\alpha_2$ together. Since the resulting
coefficients must vanish identically, one obtains the following recursive relation
\begin{align}
a_{k+1}^\mu=\frac{(-1)^{k+1} \mathcal{C}_{kp_1}(\alpha_1)-\big[k(k+1)-\mu(\mu+1)\big]a_k^\mu }{2(k+1)}.
\end{align}
The first coefficient $a_0^{\mu p_1}$ remains arbitrary and must be fixed from the initial conditions. In the spirit of
the previous approaches, we would put this coefficient equal to zero, and manoeuvre the values of the constants
$D_\mu^{(1)}(\alpha_1)$ and $D_\mu^{(2)}(\alpha_1)$ in order to meet the initial conditions. However, because of the
striking simplicity of the formula (\ref{wmuexp}), it becomes attractive to set \emph{both} of the constants equal to
zero and then use $a_0^{\mu p_1}$ to meet the initial conditions. The derivation of the analytical formula for 
$a_0^{\mu p_1}$ is presented in Supplemental Material \cite{supplement}.

We have to stress that the presented asymptotic expansion of $W_\mu^0(p_1,p_2,\alpha_1,\alpha_2)$ is valid only when
$\alpha_2$ is large and when $\alpha_2 > \alpha_1$. This happens because of the properties of the adopted series
expansion of the inhomogeneous term, Eqs. (\ref{aplarge1}) and (\ref{aplarge2}). Additionally, to assert a rapid
convergence of the series (\ref{wmuexp}), the values of $\alpha_1$ and $\alpha_2$ need to be largely spaced. Simple
numerical tests showed that the difference around 20 is a safe minimum, at least for small or moderate values of
$\alpha_1$. Of course, the larger the difference is, the faster the series (\ref{wmuexp}) converges.

The above requirements may be considered to be a huge limitation of the presented procedure. However, let us note that
the exchange integrals with both $\alpha_1$ and $\alpha_2$ large tend to be very small. As a result, they would be
probably neglected by the Schwarz inequality or a similar screening method. Therefore, a
majority of the non-negligible integrals with very large $\alpha_2$ has a significantly lower value of $\alpha_1$ (or
\emph{vice versa}), so that they fall into the regime where the asymptotic method is well-suited.

In Table \ref{table5} we present results of the calculations with our asymptotic method, compared with the ``exact'' 
values
calculated
from the analytic expression in the extended precision arithmetic. The higher values of $p_2$ in the integrals are
obtained by a consecutive differentiation of the final formula (\ref{wmuexp}) with respect to $\alpha_2$. This
differentiation is elementary, since the coefficients $a_k^{\mu p_1}$ do not depend on $\alpha_2$. It follows from Table
\ref{table5} that the results obtained with the asymptotic expansion are accurate, if only $\alpha_1$ and
$\alpha_2$ are sufficiently spaced and $\alpha_2$ is large [the roles of $\alpha_1$ and $\alpha_2$ can be interchanged
due to the symmetry relation (\ref{wmusymm})]. The convergence is also rapid in this case and at most few tens of terms
suffices to achieve the desired threshold. This results confirm the validity of the proposed asymptotic expansion, Eq.
(\ref{wmuexp}).

\begin{table*}
\caption{Exemplary calculations of the $W_\mu^0(p_1,0,\alpha_1,\alpha_2)$ functions for a few representative values of
$\alpha_1$ and $\alpha_2$. \emph{Exact} denotes values calculated using Eq. (\ref{wmugen}) in the extended arithmetic
precision of 32 significant digits with the \textsc{Mathematica} package (all digits shown are correct).
\emph{Asymptotic expansion} column shows results of calculations with Eq. (\ref{wmuexp}) in the double precision
arithmetic. \emph{Convergence} denotes a number of terms in Eq. (\ref{wmuexp}) required to converge the summation to the
maximal possible accuracy. The symbol $[k]$ denotes the powers of 10, 10$^k$.}
\begin{ruledtabular}
\begin{tabular}{cccc}
\label{table5}
$\mu$ & exact & asymptotic expansion & convergence \\
\hline\\[-2.1ex]
\multicolumn{4}{c}{ $\alpha_1=5.0$, $\alpha_2=18.0$, $p_1=0$ } \\
\hline\\[-2.1ex]
0  & 1.55 752 619 710 528 [$-12$] & 1.55 752 619 710 358 [$-12$] & 25 \\
5  & 1.20 597 911 134 310 [$-13$] & 1.20 597 911 134 129 [$-13$] & 27 \\
10 & 3.78 964 912 679 376 [$-14$] & 3.78 964 912 679 764 [$-14$] & 29 \\
15 & 1.79 942 552 611 878 [$-14$] & 1.79 942 552 612 880 [$-14$] & 31 \\
20 & 1.04 235 547 483 236 [$-14$] & 1.04 235 547 484 335 [$-14$] & 33 \\
25 & 6.77 982 765 501 726 [$-15$] & 6.77 982 765 505 423 [$-15$] & 35 \\
\hline\\[-2.1ex]
\multicolumn{4}{c}{ $\alpha_1=5.0$, $\alpha_2=18.0$, $p_1=5$ } \\
\hline\\[-2.1ex]
0  & 3.98 698 547 222 427 [$-12$] & 3.98 698 547 222 507 [$-12$] & 27 \\
5  & 1.78 103 428 717 980 [$-13$] & 1.78 103 428 718 353 [$-13$] & 29 \\
10 & 5.05 716 132 622 295 [$-14$] & 5.05 716 132 622 680 [$-14$] & 31 \\
15 & 2.33 281 860 879 113 [$-14$] & 2.33 281 860 879 422 [$-14$] & 33 \\
20 & 1.33 579 360 806 489 [$-14$] & 1.33 579 360 807 802 [$-14$] & 34 \\
25 & 8.63 919 442 650 389 [$-15$] & 8.63 919 442 656 241 [$-15$] & 37 \\
\hline\\[-2.1ex]
\multicolumn{4}{c}{ $\alpha_1=8.0$, $\alpha_2=25.0$, $p_1=0$ } \\
\hline\\[-2.1ex]
0  & 3.14 843 080 402 671 [$-46$] & 3.14 843 080 402 670 [$-46$] & 10 \\
5  & 3.14 843 080 402 671 [$-46$] & 3.14 843 080 402 670 [$-46$] & 10 \\
10 & 1.61 980 042 035 663 [$-46$] & 1.61 980 042 035 663 [$-46$] & 12 \\
15 & 9.77 378 855 083 714 [$-47$] & 9.77 378 855 083 711 [$-47$] & 15 \\
20 & 6.46 965 882 608 025 [$-47$] & 6.46 965 882 608 030 [$-47$] & 19 \\
25 & 4.56 383 003 051 265 [$-47$] & 4.56 383 003 051 411 [$-47$] & 22 \\
\hline\\[-2.1ex]
\multicolumn{4}{c}{ $\alpha_1=8.0$, $\alpha_2=25.0$, $p_1=5$ } \\
\hline\\[-2.1ex]
0  & 3.14 843 080 402 671 [$-46$] & 3.14 843 080 402 670 [$-46$] & 10 \\
5  & 3.14 843 080 402 671 [$-46$] & 3.14 843 080 402 670 [$-46$] & 10 \\
10 & 1.61 980 042 035 663 [$-46$] & 1.61 980 042 035 663 [$-46$] & 12 \\
15 & 9.77 378 855 083 714 [$-47$] & 9.77 378 855 083 711 [$-47$] & 15 \\
20 & 6.46 965 882 608 025 [$-47$] & 6.46 965 882 608 030 [$-47$] & 19 \\
25 & 4.56 383 003 051 265 [$-47$] & 4.56 383 003 051 411 [$-47$] & 22 \\[0.2ex]
\end{tabular}
\end{ruledtabular}
\end{table*}

\subsection{Final remarks on the analytic methods of calculation of the $W_\mu^0(p_1,p_2,\alpha_1,\alpha_2)$ functions}
\label{subsec:remarks}

After presenting the working formulae, let us briefly summarise the advances reported in this section. We have derived
the differential equation for $W_\mu^0(p_1,p_2,\alpha_1,\alpha_2)$ functions with respect to the nonlinear parameter
$\alpha_2$. Upon this differential equation, two important new methods of calculations have been built. The first
one is aimed at the small $\alpha_i$ regime, where the analytical expression, Eq. (\ref{wmugen}), is numerically
unstable. The second method provides an efficient and reliable method to calculate $W_\mu^0(p_1,p_2,\alpha_1,\alpha_2)$
for the asymptotically large values of $\alpha_2$ or $\alpha_1$. Each of the methods of calculation has its own
drawbacks and limitations, which have been stressed earlier. Therefore, we have to investigate how these methods can be
combined in order to produce a general algorithm. We also need to carefully check if the available methods cover
the whole area of interest.

Figure I presents the first quarter of the $(\alpha_1,\alpha_2)$ plane which corresponds to all possible combinations
of the physically relevant integrals. We divide this plane into several non-overlapping regions in which different
methods of computation can be used in a stable and efficient manner. Generally speaking, we introduce four numerical
parameters $\lambda_1$, $\delta_1$, $\lambda_2$, and $\delta_2$ which control switching between algorithms:
\begin{itemize}
\item when the difference $|\alpha_1-\alpha_2|$ is larger than $\delta_1$ and either $\alpha_1$ or $\alpha_2$ is
smaller than $\lambda_1$, the small $\alpha$ expansion presented in Subsection \ref{subsec:wmusmall} is used,
\item when both $\alpha_1$ and $\alpha_2$ are larger than $\lambda_2$ and the difference $|\alpha_1-\alpha_2|$ is larger
than $\delta_2$, the asymptotic formulae described in Subsection \ref{subsec:wmularge} need to be used,
\item when both $\alpha_1$ and $\alpha_2$ are larger than $\lambda_1$ and the requirements of the asymptotic
expansion are not met, the analytical expression (\ref{wmugen}) is used.
\end{itemize}
The separation described above is depicted graphically on Figure I. This is a direct result of the symmetry relation,
Eq. (\ref{wmusymm}).
The actual values of the parameters $\lambda_1$, $\delta_1$, $\lambda_2$, and $\delta_2$ need to be chosen on the basis
of the numerical experiments. Our current estimate for ``the best'' values is $\lambda_1\approx3$,
$\lambda_2\approx25$, $\delta_1\approx2$, and $\delta_2\approx20$. This choice is purely ``empirical'' and we are rather
conservative in this respect. These values might change after gaining more numerical experience and observing the
performance of the production code. One can even imagine that these values can slightly be modified from one basis set
to another to match their specific requirements. Nonetheless, we believe that the values suggested by us are close to
optimal.

From this brief study of the introduced regions one slightly afflictive conclusion can be drawn. There exists a 
region which is not covered by any of the analytical methods presently available. All methods are either numerically
unstable or just invalid in the region where both $\alpha_1$ and $\alpha_2$ are smaller than $\lambda_1$ and the
difference $|\alpha_1-\alpha_2|$ is smaller than $\delta_1$. However, this region is very small compared to the initial
vast area of ``no-man's land'' before our methods were introduced. Numerical tests show that for typical basis sets and
reasonable values of the internuclear separation, at most a few percent of the integrals fall in the problematic regime.
Practically, it means that this region can be treated using a more computationally expensive, and possibly a purely
numerical method without a significant overhead. The next section of the paper is devoted entirely to the development of
the ``last resort'' numerical integration scheme which completes the theory.

\begin{figure}
\caption{The $(\alpha_1,\alpha_2)$ plane which corresponds to all possible combinations
of the physically relevant integrals. The plane is divided into several regions in which different methods
of computation of $W_\mu^0(p_1,p_2,\alpha_1,\alpha_2)$ are used. See Section \ref{subsec:remarks} for the discussion
and comments. }
\label{fig1s}
\includegraphics[scale=0.50]{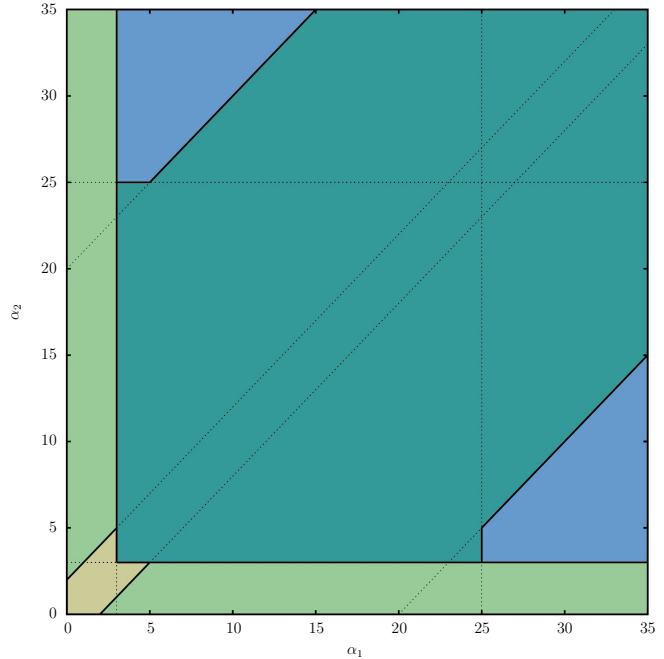}
\end{figure}

\section{The ``last resort'' numerical integration approach}

As mentioned above, to make the presented theory complete, we need a method which is reliable in the region where both
$\alpha_1$ and $\alpha_2$ are smaller than $\lambda_1$ and the difference $|\alpha_1-\alpha_2|$ is smaller than
$\delta_1$. Since this region is tiny and a small number of integrals fall within it, a more expensive method can be
used there. We propose to overcome this last obstacle by using a numerical integration. We consider it to be a
temporary remedy, useful until a new analytical approach appear.

A direct numerical integration of Eq. (\ref{smallw}) is rather daunting. Despite the apparent simplicity of the
integrand, a two dimensional quadrature rule has to be used. Such an approach has been pursued in the literature 
\cite{kolos60a,kolos60b} but the
resulting algorithms are typically very slow. We would like to adopt another line of attack. The inner integral in
(\ref{wmugen}) is worked out analytically by using a general recursive process, and the outer one-dimensional
integration is carried out numerically.

Let us first introduce a slightly more general integral class
\begin{align}
\label{wmunu}
\begin{split}
w_{\mu\nu}^\sigma(p_1,p_2,\alpha_1,\alpha_2)&=\int_1^\infty d\xi_1\, Q_\mu^{\sigma}(\xi_1)\,
(\xi_1^2-1)^{\sigma/2}\xi^{p_1}e^{-\alpha_1\xi_1}\\
&\times\int_1^{\xi_1}d\xi_2\, P_\mu^{\sigma}(\xi_2)\,
(\xi_2^2-1)^{\sigma/2}\xi^{p_2}e^{-\alpha_2\xi_2},
\end{split}
\end{align}
and
\begin{align}
\begin{split}
W_{\mu\nu}^\sigma(p_1,p_2,\alpha_1,\alpha_2)&=w_{\mu\nu}^\sigma(p_1,p_2,\alpha_1,\alpha_2)\\
&+w_{\nu\mu}^\sigma(p_2,p_1 ,\alpha_2\alpha_1),
\end{split}
\end{align}
so that integrals with $\mu=\nu$ correspond to the desired values and off-diagonal terms serve
as auxiliary quantities. It becomes obvious that the values of $p_1$ and $p_2$ can easily be increased by means of the
recurrence relation
\begin{align}
\begin{split}
&W_{\mu\nu}^\sigma(p_1+1,p_2,\alpha_1,\alpha_2)=\\
&\frac{\mu+\sigma+1}{2\mu+1}W_{\mu+1,\nu}^\sigma(p_1,p_2,\alpha_1,
\alpha_2)\\
&+\frac{\mu-\sigma}{2\mu+1}W_{\mu-1,\nu}^\sigma(p_1+1,p_2,\alpha_1,\alpha_2),
\end{split}
\end{align}
and from an analogous one for the parameter $p_2$. Note, that the above recursion is not self-starting, but initial 
values can be
obtained by the procedure similar as used by Harris \cite{harris02} which is sufficiently numerically stable. When
$p_1$ and $p_2$ are increased in this way, it remains to calculate the integrals with $p_1$=$p_2$=$0$. To proceed
further we introduce the following function which is, in substance, the inner integral in Eq. (\ref{wmunu}) at
$p_2=0$:
\begin{align}
\label{kinc}
\bar{k}^\sigma_\mu(x,\alpha)=\int_1^x d\xi\, P_\mu^\sigma(\xi)(\xi^2-1)^{\sigma/2}\,e^{-\alpha\xi}.
\end{align}
Note, that the above integrals obey the recursion relation (\ref{kmusigma}). Herein, we shall use this recursion in
a somehow different direction
\begin{align}
\label{reck}
(2\mu+1)\bar{k}^{\sigma+1}_\mu(x,\alpha) + \bar{k}^\sigma_{\mu+1}(x,\alpha) =
\bar{k}^\sigma_{\mu-1}(x,\alpha),
\end{align}
which can used to build all values with $\mu\leq \sigma$ starting with integrals with $\mu=\sigma$ and $\mu=\sigma-1$.
To evaluate these starting values let us recall the following explicit expressions for the Legendre functions:
\begin{align}
\label{pmumu}
&P_\mu^\mu(\xi)=\frac{(2\mu)!}{2^\mu \mu!}\,(\xi^2-1)^{\mu/2},\\
\label{pmumu1}
&P_\mu^{\mu-1}(\xi)=\frac{(2\mu)!}{2^\mu \mu!}\,\xi\,(\xi^2-1)^{\frac{\mu-1}{2}}.
\end{align}
Upon inserting the first of the above expressions in Eq. (\ref{kinc}) and applying the binomial expansion to the
term $(x+1)^\mu$ one arrives at 
\begin{align}
\bar{k}^\sigma_\mu(x,\alpha)= \frac{(2\mu)!}{2^\mu \mu!}\sum_{k=0}^\mu { \mu \choose k}
\int_1^x d\xi\,(\xi-1)^\mu \xi^k\,e^{-\alpha\xi}.
\end{align}
Next, by applying the substitution $t=(\xi-1)/(x-1)$, expanding another term containing $t+1$ with the help of the 
binomial
theorem, changing the order of summation and writing the result in terms of the $a_n$ function, Eq. (12) in Paper I, we
find
\begin{align}
\begin{split}
\label{kmumu}
\bar{k}^\mu_\mu(x,\alpha)&= e^{-\alpha} \frac{(2\mu)!}{2^\mu \mu!}(x-1)^{\mu+1}\sum_{l=0}^\mu { \mu \choose
l}\\
&\times (x-1)^l a_{\mu+l}\left[\alpha(x-1)\right]\sum_{k=l}^\mu { \mu \choose k}.
\end{split}
\end{align}
Note, that an important feature of the above expression is that no loss of digits during computation is possible, all
terms included in the double sum are positive and the functions $a_n$ can be calculated with a strictly controlled 
precision by
using the Miller algorithm \cite{gautschi67}. The second quantity necessary to initiate the recursive process,
$\bar{k}^{\mu-1}_\mu(x,\alpha)$, is evaluated by using a very similar expression which can be derived starting with Eq.
(\ref{pmumu1}). Since the derivation follows exactly the same pattern with only minor differences we do not present it
here. Let us conclude that computation of $\bar{k}^\sigma_\mu(x,\alpha)$ from Eq. (\ref{kmumu}), its
counterpart for $\sigma=\mu-1$ and recursion relation (\ref{reck}) is free of any digital erosion and virtually
guaranteed to give the machine precision in the result.

The final step of the method presented in this subsection is a numerical integration over the variable $\xi_1$.
Thus, the integral $W_{\mu\nu}^\sigma(0,0,\alpha_1,\alpha_2)$ is approximated as a finite sum
\begin{align}
\label{wwint}
\begin{split}
&W_{\mu\nu}^\sigma(0,0,\alpha_1,\alpha_2)\approx
\sum_k w_k\,(x_k^2-1)^{\sigma/2} \\
&\Big[
Q_\mu^\sigma(x_k)e^{-\alpha_1 x_k}
\bar{k}_\nu(x_k,\alpha_2)+
Q_\nu^\sigma(x_k) e^{-\alpha_2 x_k}
\bar{k}_\nu(x_k,\alpha_1)
\Big],
\end{split}
\end{align}
where $x_k$ and $w_k$ are the nodes and weights of a numerical integration rule. The other nonelementary quantities
entering Eq. (\ref{wwint}) are the scaled Legendre functions of the second kind, $Q_\mu^\sigma(x)(x^2-1)^{\sigma/2}$.
Their evaluation has been discussed many times in the literature and it seems that they are best computed by downward
recursion in $\mu$ followed by upward recursion in $\sigma$, see for instance Refs. \cite{gil98,olver83,schneider10}. 
Another troublesome aspect is
choice of the numerical quadrature. The integrand, Eq. (\ref{wwint}), is resistant to numerical
integration and the conventional choice of the Gaussian-type quadratures requires a large number of nodes to match the 
prescribed accuracy requirements. There are two
reasons for such a slow convergence with respect to the size of the quadrature. Firstly, the integrand is sharply peaked
around its maximum, especially for large $\mu$/$\nu$, and then vanishes very quickly (exponentially). Previous 
investigators also encountered this problem and proposed the so-called M\"{o}bius transformation \cite{lopez94} which 
makes the integrand more smooth and
well-behaved. This partial solution can be applied here straightforwardly. The second problem are the logarithmic 
singularities present in $Q_\mu(\xi)$ around $\xi=1$. These singularities are of course
integrable, but pose a considerable difficulty for the standard Gaussian quadratures with nonsingular weight 
functions. However, the so-called extended Gaussian quadratures are available which
are designed to integrate functions of polynomial-logarithmic type and their performance is greatly improved 
compared to the standard schemes. Recently,
accurate extended Gaussian quadratures with large number of nodes have been reported along with a general 
algorithm for computation of weights and abscissae (see Refs. \cite{karlin66,ma96,pachucki14} and references therein).

\section{Conclusions}

In this paper, which constitutes the second part of the series, we considered the problem of efficient and accurate
calculations of the two-centre exchange integrals over Slater-type orbitals. The main advancement presented here is the
derivation of the differential equations for two the most important basic quantities, the $L_\mu(p,\alpha)$ and
$W_\mu^0(p_1,p_2,\alpha_1,\alpha_2)$ functions. The obtained differential equations are subsequently used to arrive at
the series expansions for these basic functions. Series expansions for the small values of the parameters $\alpha_i$ are
used to supplement the available analytic methods in situations where the digital erosion observed in the calculations
becomes overwhelming. Asymptotic expansions for large values of $\alpha_i$ serve as a cheap alternative for the analytic
expressions and are useful for further numerical or mathematical analysis. We have also considered numerical integration
as an alternative in a small region where all analytic methods are not sufficiently accurate. All the available methods
were combined in order to produce a general algorithm which allows an accurate calculation of the basic integrals within
the whole region of practical interest.

Let us also note here that in the future much may be extended from the present work. The
differential equation (\ref{bigwdiff}), due to its mathematical simplicity and compactness, offers an encouraging
starting point for more advanced developments. Progress towards new expressions which remove the necessity to use
the numerical integration is definitely welcomed. On the other hand, a completely different direction of
the advancement can be pursued. An example could be derivation of the large $\mu$ or $\sigma$ expansions of the $L$ and
$W$ functions. Since the large $\mu$/$\sigma$ expansions of the solutions to the homogeneous differential equation
(\ref{bigwdiff}) are well-known, and the inhomogeneity does not depend on $\mu$, such efforts might likely succeed. 

The advances presented here and in the previous paper allow us to compute all molecular integrals required for the
state-of-the-art \emph{ab initio} calculations on the diatomic molecules with a reasonable speed and sufficient
accuracy. This allows us to launch an assault on the problem of bonding between two beryllium atoms. The third, and
final, paper of the series is entirely devoted to the case study of the beryllium dimer.

\begin{acknowledgments}
This work was supported by the Polish Ministry of Science and Higher Education, grant NN204 182840. ML acknowledges the
Polish Ministry of Science and Higher Education for the support through the project \textit{``Diamentowy Grant''},
number DI2011 012041. RM was supported by the Foundation for Polish Science through the \textit{``Mistrz''} program. We
would like to thank Bogumi\l~Jeziorski for fruitful discussions, reading and commenting on the manuscript, and Krzysztof
Ko\'{s}\'{c} for assistance in preparation of the manuscript.
\end{acknowledgments}

\end{document}

\title{{\Large Supplemental Material.} \\[0.5em] \noindent 
Calculation of two-centre two-electron integrals over Slater-type orbitals revisited. \\II.
Neumann expansion of the exchange integrals }

\author{\sc Micha\l\ Lesiuk\footnote[1]{e-mail: lesiuk@tiger.chem.uw.edu.pl} 
and Robert Moszynski}
\affiliation{\sl Faculty of Chemistry, University of Warsaw\\
Pasteura 1, 02-093 Warsaw, Poland}
\date{\today}
%\pacs{31.15.vn, 03.65.Ge, 02.30.Gp, 02.30.Hq}

\begin{abstract}
The present document serves as a supplemental material for the publication \emph{Calculation of two-centre
two-electron integrals over Slater-type orbitals revisited. II. Neumann expansion of the exchange integrals}. It
contains data or derivations which are of some importance but are not necessary for an overall understanding of the
manuscript. However, the material presented here will be useful for a reader who wishes to repeat all of our derivations
in details. Researchers who wish to repeat some of the calculations independently may also benefit from these
additional data.
\end{abstract}

\maketitle

\section{Supplemental material for Paper II}
\subsection{Analytical expression for $C_\mu^{(1)}$ constant in the small $\alpha$ expansion of $L_\mu^0(0,\alpha)$}

The simplest way to find an analytical expression for $C_\mu^{(1)}$ is to compare our equations (29), (30) 
with the one proposed by Harris, Eqs. (38) and (39) (from Paper II). It is obvious that in the small $\alpha$ limit both
expressions must lead to an exactly the same expansion. It is also possible to find $C_\mu^{(1)}$ by using only the
formula (12) from Paper II but the derivation is much more tedious and too long for the purposes of this paper.

Let us begin by differentiating Eqs. (29), (30) $\mu$ times with respect to $\alpha$. Next, by taking the limit
$\alpha\rightarrow0$ one obtains:
\begin{align}
\label{const1}
L_\mu^0(\mu,\alpha) = C_\mu^{(1)}\frac{ (-1)^\mu \mu!}{(2\mu+1)!!}
+(-1)^\mu d_\mu^\mu \log(2\alpha)
+(-1)^\mu \mu! H_\mu d_\mu^\mu + \mathcal{O}(\alpha)
\end{align}
where the elementary identity $i_\mu^0(\mu,0)=\frac{\mu!}{(2\mu+1)!!}$ has been used. $H_\mu$
denotes the harmonic number ($n$-th harmonic number is the sum of the reciprocals of the first $n$ natural numbers). In
the above expression the convention $c_\mu^\mu=0$ was used, according to the discussion presented in the main text. One
the other hand, one can take the same limit in Eq. (38) from Paper II for $L_\mu^0(\mu,\alpha)$. Under these
circumstances, the first term in Eq. (38) vanishes and one is left with:
\begin{align}
L_\mu^0(\mu,\alpha) =-\gamma_E \frac{\mu!}{(2\mu+1)!!}-\frac{\mu!}{(2\mu+1)!!}
\log(\alpha)+\mathcal{M}_0^{\mu0}(\mu)+ \mathcal{O}(\alpha).
\end{align}
In the derivation of the above expression the small $\alpha$ expansion of $E_1$ is useful:
$E_1(\alpha)=-\gamma_E-\log(\alpha)+\mathcal{O}(\alpha)$. Expression for the coefficient $\mathcal{M}_0^{\mu0}(\mu)$
becomes:
\begin{align}
\mathcal{M}_0^{\mu0}(\mu)=-\frac{\mu!}{(2\mu+1)!!}\log(2)+\widetilde{\mathcal{M}}_0^{\mu0}(\mu),
\end{align}
with
\begin{align}
\widetilde{\mathcal{M}}_0^{\mu0}(\mu) &= \sum_{j=0}^{\mu/2}\frac{(-1)^j(2\mu-2j-1)!!}{(\mu-2j)!(2j)!!(2\mu-2j+1)}
\sum_{k=0}^{\mu-j}\frac{1}{2k+1},
\end{align}
and it does not seem to simplify beyond that. Finally, we obtain the following expression:
\begin{align}
\label{const2}
L_\mu^0(\mu,\alpha) =-\gamma_E
\frac{\mu!}{(2\mu+1)!!}-\frac{\mu!}{(2\mu+1)!!}
\log(2\alpha)+\widetilde{\mathcal{M}}_0^{\mu0}(\mu)+ \mathcal{O}(\alpha).
\end{align}
Since the formulae (\ref{const1}) and (\ref{const2}) have to be identically the same, we can equate them. By comparing
the quantities which are proportional to the logarithmic terms one obtains an equation for $d_\mu^\mu$:
\begin{align}
-\frac{\mu!}{(2\mu+1)!!}=(-1)^\mu\mu!d_\mu^\mu,
\end{align}
and for the leftover
\begin{align}
(-1)^\mu C_\mu^{(1)}\frac{\mu!}{(2\mu+1)!!}+(-1)^\mu\mu!H_\mu d_\mu^\mu=
-\gamma_E +\widetilde{\mathcal{M}}_0^{\mu0}(\mu).
\end{align}
By solving the first of the above equations for $d_\mu^\mu$ and inserting the result back into the second equation one
finally arrives at the desired analytical expression for $C_\mu^{(1)}$:
\begin{align}
\label{cmu1}
C_\mu^{(1)} = (-1)^\mu \bigg[ H_\mu-\gamma_E+ \frac{(2\mu+1)!!}{\mu!}\widetilde{\mathcal{M}}_0^{\mu0}(\mu) \bigg].
\end{align}
For practical use, the values of $C_\mu^{(1)}$ can simply be tabulated and included in the production program, as
discussed in the main text.

\subsection{Analytical expression for $D_\mu^{(2)}$ constant in the large $\alpha$ expansion of $L_\mu^0(0,\alpha)$}

Starting with Eqs. (40) and (41) from Paper II, we multiply both sides by $\alpha e^\alpha$ and take the limit
of large $\alpha$. Taking advantage of the asymptotic formula
\begin{align}
\label{asymk}
k_\mu(\alpha) = e^{-\alpha} \left[ \frac{1}{\alpha}+\mathcal{O}\left(\frac{1}{\alpha^2}\right)\right],
\end{align}
one finds that as $\alpha\rightarrow\infty$
\begin{align}
\label{asymlc}
\alpha e^\alpha L_\mu(\alpha)=D_\mu^{(2)}+\frac{1}{2}\log(2\alpha)+\mathcal{O}\left(\frac{1}{\alpha}\right),
\end{align}
since $b_1^\mu=1/2$ and $a_1^\mu=0$. Considering now Eq. (15) from Paper II and making use of
the asymptotic formulae
\begin{align}
A_p(\alpha)=e^{-\alpha} \left[ \frac{1}{\alpha}+\mathcal{O}\left(\frac{1}{\alpha^2}\right)\right],\;\;\;
E_1(2\alpha)=e^{-2\alpha} \left[ \frac{1}{2\alpha}+\mathcal{O}\left(\frac{1}{\alpha^2}\right)\right],
\end{align}
it becomes elementary to derive
\begin{align}
\alpha e^\alpha L_0^0(p,\alpha) =
\frac{1}{2}\left[\gamma_E+\log(2\alpha)\right]+\mathcal{O}\left(\frac{1}{\alpha}\right).
\end{align}
By inserting the above expression into Eq. (12) from Paper II and by using the asymptotic formula for $A_p$ functions
one
arrives at
\begin{align}
\alpha e^\alpha L_\mu(\alpha)=\frac{1}{2}\sum_s^{\mu} \mathcal{A}_s^{\mu0}\left[\gamma_E+\log(2\alpha)\right]
+\sum_s^{\mu-1}\mathcal{B}_s^{\mu0} \cdot 1+\mathcal{O}\left(\frac{1}{\alpha}\right).
\end{align}
It is rather simple to show that 
\begin{align}
\sum_s^{\mu} \mathcal{A}_s^{\mu0}&=1,\;\;\;\sum_s^{\mu-1}\mathcal{B}_s^{\mu0}=-H_\mu,
\end{align}
and finally
\begin{align}
\alpha e^\alpha
L_\mu(\alpha)=\frac{1}{2}\left[\gamma_E+\log(2\alpha)\right]-H_\mu+\mathcal{O}\left(\frac{1}{\alpha}\right).
\end{align}
By comparing the above expression with the initial formula (\ref{asymlc}) we arrive at the desired expression
$D_\mu^{(2)}=\frac{1}{2}\gamma_E-H_\mu$,
which is trivial to calculate and completes the asymptotic theory presented in Subsection IV D of Paper II.

\subsection{Analytical expression for the $C_\mu^{(1)}(\alpha_1)$ constant in the small $\alpha_2$ expansion of
$W_\mu^0(p_1,p_2,\alpha_1,\alpha_2)$}

Let us begin with expressions (62) and (63) from Paper II. Let us differentiate these formulae $\mu$ times with respect
to $\alpha_2$. Since $C_{\mu p_1}^{(2)}(\alpha_1)$ and $a_\mu^{\mu p_1}$ have already been set to zero (see the main
text for the discussion) one readily obtains the following expression in the limit $\alpha_2 \rightarrow 0^+$:
\begin{align}
\label{wmu0a}
W_\mu(p_1,\mu,\alpha_1,\alpha_2) = 
(-1)^\mu \mu! \left[
\frac{C_{\mu p_1}^{(1)}(\alpha_1)}{(2\mu+1)!!}
+e^{-\alpha_1}H_\mu\,b_\mu^{\mu p_1}+
e^{-\alpha_1}\,b_\mu^{\mu p_1} \log(2\alpha_2)
\right]+\mathcal{O}\left(\alpha_2\right).
\end{align}
Let us now consider the limit $\alpha_2 \rightarrow 0^+$ in Eq. (5) from Paper II with $p_2=\mu$ and $\sigma=0$. The 
first term on the right hand side is regular at $\alpha_2=0$ but the second term needs to be rewritten in the following 
way:
\begin{align}
\begin{split}
&w_\mu(\mu,p_1,\alpha_2,\alpha_1)=\int_1^\infty d\xi_1 Q_\mu(\xi_1)\xi_1^\mu e^{-\alpha_2\xi_1}
\int_1^{\xi_1} d\xi_2 P_\mu(\xi_2)\xi_2^{p_1}e^{-\alpha_1\xi_2}=
\int_1^\infty d\xi_1 Q_\mu(\xi_1)\xi_1^\mu e^{-\alpha_2\xi_1}
\int_1^{\xi_1} d\xi_2 P_\mu(\xi_2)\xi_2^{p_1}e^{-\alpha_1\xi_2}\\
-&\int_1^\infty d\xi_1 Q_\mu(\xi_1)\xi_1^\mu e^{-\alpha_2\xi_1}
\int_{\xi_1}^\infty d\xi_2 P_\mu(\xi_2)\xi_2^{p_1}e^{-\alpha_1\xi_2}=
k_\mu(p_1,\alpha_1)L_\mu(\mu,\alpha_2)-\bar{w}_\mu(\mu,p_1,\alpha_2,\alpha_1).
\end{split}
\end{align}
Note, that now the first term in the above expression contains the logarithmic singularity at $\alpha_2=0$ but the 
second term is regular. Additionally, we can now make use of Eq. (\ref{const2}) to obtain the small $\alpha_2$ formula:
\begin{align}
\begin{split}
w_\mu(\mu,p_1,\alpha_2,\alpha_1)=
k_\mu(p_1,\alpha_1)\left[ -\gamma_E
\frac{\mu!}{(2\mu+1)!!}-\frac{\mu!}{(2\mu+1)!!}
\log(2\alpha_2)+\widetilde{\mathcal{M}}_0^{\mu0}(\mu) \right]-\bar{w}_\mu(\mu,p_1,0,\alpha_1)+ \mathcal{O}(\alpha_2),
\end{split}
\end{align}
and returning to Eq. (5) from Paper II we finally have
\begin{align}
\begin{split}
\label{wmu0b}
W_\mu(p_1,\mu,\alpha_1,\alpha_2) &= 
w_\mu(p_1,\mu,\alpha_1,0)-\bar{w}_\mu(\mu,p_1,0,\alpha_1)+k_\mu(p_1,\alpha_1)\times\\
&\times\left[ -\gamma_E
\frac{\mu!}{(2\mu+1)!!}-\frac{\mu!}{(2\mu+1)!!}
\log(2\alpha_2)+\widetilde{\mathcal{M}}_0^{\mu0}(\mu) \right]+ \mathcal{O}(\alpha_2),
\end{split}
\end{align}
It is now possible to compare the expressions (\ref{wmu0a}) and (\ref{wmu0b}). Terms proportional to the logarithm give 
rise to the following equality
\begin{align}
(-1)^\mu \mu! e^{-\alpha_1}\,b_\mu^{\mu p_1} = -\frac{\mu!}{(2\mu+1)!!} k_\mu(p_1,\alpha_1),
\end{align}
and the leftover 
\begin{align}
\label{ccc}
\begin{split}
(-1)^\mu \mu! \left[
\frac{C_{\mu p_1}^{(1)}(\alpha_1)}{(2\mu+1)!!}
+e^{-\alpha_1}H_\mu\,b_\mu^{\mu p_1}\right]&=
w_\mu(p_1,\mu,\alpha_1,0)-\bar{w}_\mu(\mu,p_1,0,\alpha_1)\\
&+k_\mu(p_1,\alpha_1)\left[ -\gamma_E \frac{\mu!}{(2\mu+1)!!}+\widetilde{\mathcal{M}}_0^{\mu0}(\mu) \right].
\end{split}
\end{align}
The first of the above expressions can be solved for $b_\mu^{\mu p_1}$ and the result is inserted in the second one. 
After some rearrangements the final result reads:
\begin{align}
\frac{(-1)^\mu \mu!}{(2\mu+1)!!}C_{\mu p_1}^{(1)}(\alpha_1)=
w_\mu(p_1,\mu,\alpha_1,0)-\bar{w}_\mu(\mu,p_1,0,\alpha_1)
+k_\mu(p_1,\alpha_1)\left[ \left(H_\mu-\gamma_E\right) \frac{\mu!}{(2\mu+1)!!}+\widetilde{\mathcal{M}}_0^{\mu0}(\mu) 
\right].
\end{align}
The above expression is of little use if there is no simple way to calculate the values of $w_\mu(p_1,\mu,\alpha_1,0)$ 
and $\bar{w}_\mu(\mu,p_1,0,\alpha_1)$. In the following we show that those two integrals can be simply expressed 
through the $L_\mu$ functions. The simplest way to obtain the integral $w_\mu(p_1,\mu,\alpha_1,0)$ is to start with Eq. 
(54) from the 2002 paper of Harris [F. E. Harris, Int. J. Quantum Chem. \textbf{88}, 701 (2002)]. When $\alpha_2=0$ one 
obtains simply:
\begin{align}
w_\mu(p_1,\mu,\alpha_1,0) = \sum_s^\mu A_s^{\mu0}
\frac{1}{\mu+s+1}L_\mu(p_1+\mu+s+1,\alpha_1)-L_\mu(p_1,\alpha_1)\bar{i}_\mu(\mu,0),
\end{align}
where
\begin{align}
\bar{i}_\mu(\mu,0)=\int_0^1 d\xi P_\mu(\xi)\,\xi^\mu=\sum_s^\mu A_s^{\mu0}\frac{1}{\mu+s+1}.
\end{align}
Therefore, the explicit expression is
\begin{align}
\label{ww1}
w_\mu(p_1,\mu,\alpha_1,0) = \sum_s^\mu \frac{A_s^{\mu0}}{\mu+s+1} \bigg[
L_\mu(p_1+\mu+s+1,\alpha_1)-L_\mu(p_1,\alpha_1)\bigg].
\end{align}
To calculate the second required ingredient - analytical expression for $\bar{w}_\mu(\mu,p_1,0,\alpha_1)$ one starts 
with the explicit expression
\begin{align}
\bar{w}_\mu(\mu,p_1,0,\alpha_1)=
\int_1^\infty d\xi_1\,Q_\mu(\xi_1)\,\xi_1^\mu 
\int_{\xi_1}^\infty d\xi_2\,P_\mu(\xi_2)\,\xi_2^{p_1}\,e^{-\alpha_1\xi_2}=
\sum_s^\mu A_s^{\mu0}\int_1^\infty d\xi_1\,Q_\mu(\xi_1)\,\xi_1^\mu 
\int_{\xi_1}^\infty d\xi_2\,\xi_2^{p_1+s}\,e^{-\alpha_1\xi_2},
\end{align}
and by substitution of variables $t=\xi_2/\xi_1$ in the inner integral one immediately recognises that it can 
be rewritten as
\begin{align}
\bar{w}_\mu(\mu,p_1,0,\alpha_1)=
\sum_s^\mu A_s^{\mu0}\int_1^\infty d\xi_1\,Q_\mu(\xi_1)\,\xi_1^{\mu+p_1+s+1}A_{p_1+s}(\alpha_1\xi_1).
\end{align}
Finally, when Eq. (19) from Paper I is inserted one obtains
\begin{align}
\bar{w}_\mu(\mu,p_1,0,\alpha_1)=
\sum_s^\mu A_s^{\mu0}\frac{(p_1+s)!}{\alpha_1^{p_1+s+1}}\sum_{k=0}^{p_1+s}\frac{\alpha_1^k}{k!}\,L_\mu(\mu+k,\alpha_1),
\end{align}
which formally completes the theory. Note, that Eq. (\ref{ccc}) introduces some digital erosion but it is much better 
conditioned than the original expressions of Maslen and Trefry when $\alpha_1$ is small.

\subsection{Analytical expression for the $a_k^{\mu p_1}$ constant in the large $\alpha_2$ expansion of
$W_\mu^0(p_1,p_2,\alpha_1,\alpha_2)$}

If one considers the asymptotic formula, Eq. (73) from Paper II, the following expression is elementary
\begin{align}
\lim_{\alpha_2\rightarrow\infty} e^{\alpha_1+\alpha_2}\alpha_2 W_\mu^0(p_1,0,\alpha_1,\alpha_2)=
\lim_{\alpha_2\rightarrow\infty} e^{\alpha_1+\alpha_2}\alpha_2 \mathcal{W}_\mu^{p_1,\infty}(\alpha_1,\alpha_2)=
a_0^{\mu p_1},
\end{align}
since both constants $D_\mu^{(1)}(\alpha_1)$ and $D_\mu^{(2)}(\alpha_1)$ are
\emph{a priori} set identically equal to zero for all values of the parameters. On the other hand, let us make use of
the analytical expressions for $W_\mu^0(p_1,0,\alpha_1,\alpha_2)$, Eqs. (5) and (17) from Paper II. Let us first
note, that among many terms present in this equation, only the term $L_\mu^0(p_1,\alpha_1)
k_\mu^0(0,\alpha_2)$ contributes to the above limit. This fact can be deduced straightforwardly by using the
asymptotic theory of the $L_\mu^0(p,\alpha)$ functions given earlier and Eq. (\ref{asymk}). In other words, one can show
that
\begin{align}
\lim_{\alpha_2\rightarrow\infty} e^{\alpha_1+\alpha_2}\alpha_2 
\bigg[ W_\mu^0(p_1,0,\alpha_1,\alpha_2)
- L_\mu^0(p_1,\alpha_1) k_\mu^0(0,\alpha_2) \bigg]=0,
\end{align}
for every $\mu$, $p_1$ and $\alpha_1$. Therefore
\begin{align}
\lim_{\alpha_2\rightarrow\infty} e^{\alpha_1+\alpha_2}\alpha_2 W_\mu^0(p_1,0,\alpha_1,\alpha_2)=
e^{\alpha_1} L_\mu^0(p_1,\alpha_1) \lim_{\alpha_2\rightarrow\infty}e^{\alpha_2} \alpha_2 k_\mu^0(0,\alpha_2)=
e^{\alpha_1} L_\mu^0(p_1,\alpha_1),
\end{align}
by the virtue of Eq. (\ref{asymk}) and the final formula reads
$a_0^{\mu p_1} = e^{\alpha_1} L_\mu^0(p_1,\alpha_1)$.
As one can see, in the asymptotic theory of the $W_\mu^0(p_1,p_2,\alpha_1,\alpha_2)$ functions, the necessity to
calculate $L_\mu^0(p_1,\alpha_1)$ integrals with arbitrary values of the parameters is indispensable.